\documentclass[10pt,journal,compsoc]{IEEEtran}
\ifCLASSOPTIONcompsoc
  \usepackage[nocompress]{cite}
\else
  \usepackage{cite}
\fi
\ifCLASSINFOpdf
\else
\fi

\newtheorem{theorem}{Theorem}

\newtheorem{lemma}{Lemma}

\usepackage{amsfonts}
\usepackage{graphicx}
\usepackage{amssymb}
\usepackage{hyperref}
\usepackage{bm}
\usepackage[ruled]{algorithm2e}
\usepackage{amsmath}
\usepackage{dsfont}
\usepackage{epstopdf}
\usepackage{ragged2e}
\usepackage{color}
\usepackage{verbatim}
\newtheorem{assumption}{Assumption}[section] 
\usepackage{multirow}
\usepackage{subfigure}
\usepackage{makecell}
\usepackage{hyperref}
\hypersetup{
    colorlinks=true,       
    linkcolor=black,        
    citecolor=black,       
    urlcolor=black          
}
\usepackage{bm}

\makeatletter
\def\ps@IEEEtitlepagestyle{%
	\def\@oddfoot{\mycopyrightnotice}%
	\def\@evenfoot{}%
}
\def\mycopyrightnotice{%
	{\footnotesize 
		\begin{minipage}{\textwidth}
			\centering
			\textbf{This work has been submitted to the IEEE for possible publication. Copyright may be transferred without notice, after which this version may no longer be accessible.}
		\end{minipage}
		\hfill}
	\gdef\mycopyrightnotice{}
}
\begin{document}
%
\title{FedRFQ: Prototype-Based Federated Learning with Reduced Redundancy, Minimal Failure, and Enhanced Quality}
%
%
%
%

\author{Biwei~Yan,
	  Hongliang~Zhang,
        Minghui~Xu,~\IEEEmembership{Member,~IEEE},
        Dongxiao~Yu,~\IEEEmembership{Senior~Member,~IEEE},
        Xiuzhen~Cheng,~\IEEEmembership{Fellow,~IEEE}

\IEEEcompsocitemizethanks{
	\IEEEcompsocthanksitem B. Yan, M. Xu (corresponding author), D. Yu, and X. Cheng are with the School of Computer Science and Technology, Shandong University, Qingdao 266237, China. Email: \{bwyan, mhxu, dxyu, xzcheng\}@sdu.edu.cn.
	\IEEEcompsocthanksitem H. Zhang is with the School of Computer Science and Technology, Qilu University of Technology (Shandong Academy of Sciences), Jinan, Shandong, 250353, China; E-mail: hongliangzhang2022@163.com.



}

\thanks{Manuscript received ***; revised ***}}

%
%

\markboth{Journal of \LaTeX\ Class Files,~Vol.~14, No.~8, August~2015}%
{Shell \MakeLowercase{\textit{et al.}}: Bare Demo of IEEEtran.cls for Computer Society Journals}
%



\IEEEtitleabstractindextext{%
\begin{abstract}
\justifying
Federated learning is a powerful technique that enables collaborative learning among different clients. Prototype-based federated learning is a specific approach that improves the performance of local models  under non-IID (non-Independently and Identically Distributed) settings by integrating class prototypes. However, prototype-based federated learning faces several challenges, such as prototype redundancy and prototype failure, which limit its accuracy. It is also susceptible to poisoning attacks and server malfunctions, which can degrade the prototype quality. To address these issues, we propose FedRFQ, a prototype-based \underline{fed}erated learning approach that aims to reduce \underline{r}edundancy, minimize \underline{f}ailures, and improve \underline{q}uality. FedRFQ leverages a SoftPool mechanism,  which effectively mitigates prototype redundancy and prototype failure on non-IID data. Furthermore, we introduce the BFT-detect, a BFT (Byzantine Fault Tolerance) detectable aggregation algorithm, to ensure the security of FedRFQ against poisoning attacks and server malfunctions. Finally, we conduct experiments on three different datasets, namely MNIST, FEMNIST, and CIFAR-10, and the results demonstrate that FedRFQ outperforms existing baselines in terms of accuracy when handling non-IID data.
\end{abstract}

}

\maketitle

\IEEEdisplaynontitleabstractindextext

%
\IEEEpeerreviewmaketitle

\IEEEraisesectionheading{\section{Introduction}\label{introduction}}

\IEEEPARstart{P}{rototype-based} federated learning allows heterogeneous clients to train high-quality local models based on their personalized features. In prototype-based federated learning, the exchange of class prototypes between a client and a server replaces the traditional exchange of gradients \cite{tan2022fedproto}. These prototypes are computed by taking average of the feature representations of the samples belonging to the same class \cite{tan2022fedproto,snell2017prototypical}. The use of prototypes can help federated learning overcome issues such as data heterogeneity or model heterogeneity, resulting in faster convergence and better generalization performance \cite{qiao2023prototype}. However, prototype-based federated learning still faces three challenges \textbf{[C1-C3]}.

\textbf{[C1] Prototype Redundancy.} Prototype redundancy is an urgent challenge in prototype-based federated learning \cite{li2023feature}. When the performance of a local model achieved by training with a smaller number of prototype parameters is equivalent to that obtained by training with a larger number of prototype parameters, the extra prototype parameters are considered redundant. This phenomenon is called prototype redundancy, which can lead to increased communication overhead and computing burden on the server. As far as we know, currently, there exists no prototype-based federated learning scheme specifically designed to address this issue. 

\begin{figure}[!tb]
	\centering 
	\includegraphics[scale=0.45]{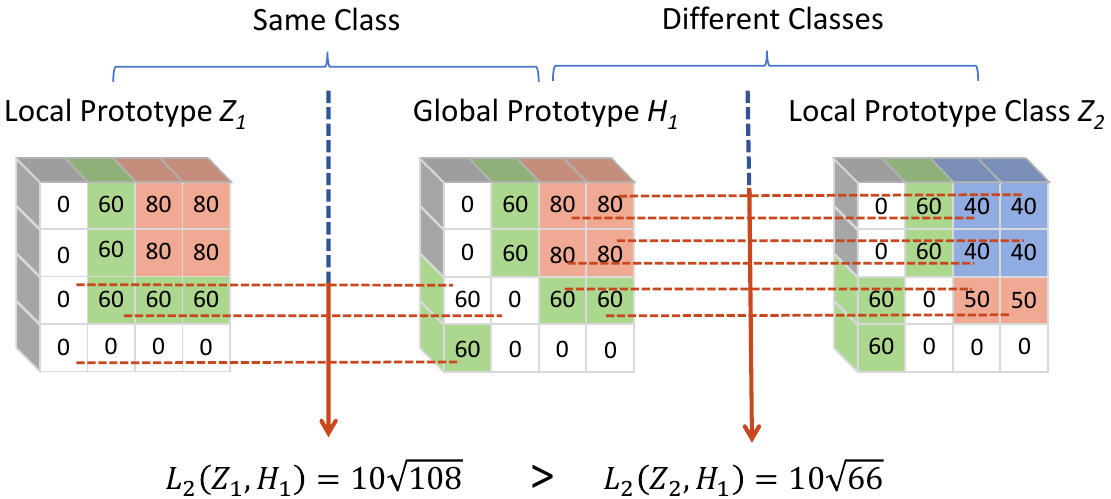}
	\caption{An example of prototype failure}
	\label{prototype}
\end{figure}

\textbf{[C2] Prototype Failure.} Prototype failure is another challenge in prototype-based federated learning. To quantify and illustrate this phenomenon, we employ the $L_2$ distance metric to measure the disparity between a local prototype and a global prototype \cite{tan2022fedproto, qiao2023prototype, qiao2023boosting, qiao2023mp}. As shown in Fig. \ref{prototype}, $L_2(Z_1, H_1)$ denotes the $L_2$ distance between the local prototype $Z_1$ and the global prototype $H_1$ of the same class, while $L_2(Z_2, H_1)$ represents the $L_2$ distance between the local prototype $Z_2$ and $H_1$, the global prototype of a different class. Ideally, $L_2(Z_1, H_1)$ should be smaller than  $L_2(Z_2, H_1)$. However, in this example, $L_2(Z_1, H_1)$ is greater than $L_2(Z_2, H_1)$. We term this phenomenon prototype failure. To gain a better understanding regarding why prototype failures can occur, we have analyzed the failures through unsupervised clustering and observed that prototypes from different classes may have similarities, making it hard to distinguish between prototypes of the same or different classes using distance-based methods.

\textbf{[C3] Low Prototype Quality.} Low-fidelity prototypes can arise due to poisoning attacks on the client side \cite{kalapaaking2023blockchain} or server-side malfunctions \cite{9615370}, resulting in compromised training outcomes and decreased accuracy. Poisoning attacks involve the manipulation of data labels by malicious entities, thereby undermining the quality of the prototypes and subsequently reducing the accuracy of prototype-based federated learning. Server malfunctions can significantly disrupt the training process, leading to failures in model aggregation and synchronization, and potentially resulting in data loss. These ramifications contribute to low prototype quality, thus adversely affecting the accuracy of prototype-based federated learning. 
 
We propose FedRFQ to address the issues of prototype redundancy, prototype failure, and low prototype quality in prototype-based federated learning. The main contributions of the paper can be summarized as follows.

\begin{itemize}
       \item We have addressed the issues of prototype redundancy and prototype failure by embedding the SoftPool pooling method in the representation layer of the local models. This approach effectively reduces redundancy and minimizes failures, thereby improving the training efficiency with non-IID (non-Independently and Identically Distributed) data in FedRFQ. To our knowledge, this is the first attempt to resolve the problems of prototype redundancy and prototype failure in prototype-based federated learning.
       
	\item We have developed a BFT (Byzantine Fault Tolerant) detectable aggregation algorithm, named BFT-detect, which ensures decentralized secure aggregation for local prototypes. This algorithm is designed to filter out low-quality prototypes and prevent server malfunctions through the consensus among servers. As a result, FedRFQ security is guaranteed and poisoning attacks are effectively resisted.
 
        \item Rigorous convergence analysis of FedRFQ in the presence of poisoning attacks and server malfunctions demonstrates its security properties. Experimental results confirm FedRFQ's effectiveness.
\end{itemize}

\textbf{Paper Organization.} The rest of this paper is organized as follows. Section \ref{related} presents the related work. In Section \ref{PRELIMINARIES}, we describe the preliminary and then introduce the model and design goals. Section \ref{FedProPool} illustrates the design details of FedRFQ. Section \ref{THEORETICAL} and Section \ref{experiments} provide the analysis and the evaluation results. Finally, we conclude the paper in Section \ref{conclusion}.

\section{Related work}\label{related}
Traditional federated learning faces the challenge of non-IID  \cite{cao2022c2s,arivazhagan2019federated,xie2021multi,collins2021exploiting}. This challenge poses difficulties in achieving convergence and the overall global model performance improvement \cite{li2019convergence}. In this section, we first introduce the most related work addressing the non-IID challenge, then overview the effort tackling the malicious behaviors in federated learning, and finally summarize what FedRFQ does to overcome the drawbacks of existing approaches. 

\subsection{Federated Learning with Non-IID}
The earlier effort to address the non-IID issue in federated learning can be summarized into three categories. The first category aims at leveraging the knowledge of global and local models from other clients to reformulate local training objectives \cite{li2020federated,li2021model,li2022federated}. FedProx \cite{li2020federated} makes use of the $L_2$ distance metric to constrain local updates, effectively dealing with the inherent heterogeneity in federated learning. Li et al. \cite{li2021model} introduced a model-contrastive federated learning paradigm, which emphasizes correcting local training through model representation similarities. Diao et al. \cite{li2022federated} proposed a partitioning strategy to cover comprehensive distribution scenarios of the non-IID data. The second category intends to calibrate the biased global model \cite{luo2021no,zhang2022fine}. In \cite{luo2021no}, Luo et al. fine-tuned the classifier by adjusting it using virtual representations sampled from an approximated Gaussian mixture model. Zhang et al.  \cite{zhang2022fine} calibrated the global model by a data-free knowledge distillation method to alleviate the performance drop caused by model aggregation. However, the approaches mentioned above still exhibit low performance when faced with significant heterogeneity in data distribution.

To further improve the performance of federated learning under non-IID, the third category, namely personalized federated learning, was proposed. Personalized federated learning can individually adjust the model training process based on the individual characteristics of the participants \cite{wu2020fedhome,10177489,9492755}. Wu et al. \cite{wu2020fedhome} introduced an innovative framework for in-home healthcare services called Edge-Cloud personalized federated learning, which involves training individual local models at home. Ye et al. \cite{10177489} designed a personalized federated multi-task learning method aiming to effectively mitigate the non-IID challenge by establishing inherent clustering patterns among clients and personalizing cluster models through the inclusion of personalized patch layers. Mills et al. \cite{9492755} integrated non-federated batch-normalization layers into deep neural networks, allowing users to fine-tune the training models according to their specific datasets and improve the accuracy of individual user models. Although these personalized federated learning methods can enhance the accuracy of model training on non-IID data, they require uploading the model parameters to the server, resulting in high communication costs.

\subsection{Prototype-based Federated Learning with Non-IID}

In order to decrease the communication overhead while addressing the data heterogeneity (non-IID) issue, prototype-based federated learning has been proposed in recent years \cite{tan2022fedproto,9643454,zhou2022fedfa,xu2023personalized,dai2023tackling,mu2023fedproc,10041939,10192272}. Tan et al. \cite{tan2022fedproto} introduced the concept of prototype as a tool to improve tolerance to heterogeneity. In their approach, clients and servers communicate prototypes instead of gradients, which improves the communication cost since the prototypes to be transmitted are much smaller in size than the gradients. Wu et al.  \cite{9643454} proposed a knowledge-sharing approach among clients, enabling the creation of client-specific representations through a prototype-wise contrastive methodology. Zhou et al. \cite{zhou2022fedfa} developed a federated learning framework called FedFA, which aligns class prototypes and calibrates classifiers across clients during local training, ensuring consistent classifiers while updating client models with shared class prototypes. Xu et al. \cite{xu2023personalized} explicitly addressed the alignment of local-global prototypes based on the global semantic knowledge to improve representation. Dai et al. \cite{dai2023tackling} proposed FedNH, a novel method that improves the performance of the local models by using uniformity and semantics in class prototypes. Mu et al. introduced the FedProc framework, a prototypical contrastive federated learning approach that treats prototypes as global knowledge to adjust drift in local training \cite{mu2023fedproc}. Zhang et al. introduced a federated learning strategy, FedPM, which collectively learns a DCNN model using the remote sensing data \cite{10041939}. Yu et al. \cite{10192272} constructed an improved instance normalization module to enhance the discrimination and generalization ability of the local model. They also introduced a contrastive loss based on prototypes to align classes in the embedding space.

These prototype-based methods do show potential in addressing the non-IID problem and improving communication cost. However, they still exhibit inherent shortcomings, such as prototype redundancy and prototype failure, which can lead to decreased accuracy and efficiency.

\subsection{Federated Learning with Malicious Attacks}
Federated learning is susceptible to malicious activities such as poisoning attacks, prompting researchers to propose various schemes aimed at identifying and resisting such attacks \cite{9347812,chen2021towards,9272656,liu2021privacy,10184496,10136231,kalapaaking2023blockchain}. Kalapaaking et al. \cite{kalapaaking2023blockchain} presented a federated learning framework to resist poisoning attacks. The framework utilizes secure multi-party computation to verify the models and eliminate the poisoned ones, while preserving the privacy of the local models. Liu et al. \cite{liu2021privacy} introduced a privacy-enhanced federated learning framework, which utilizes the Pearson correlation coefficient to detect malicious gradients and employs adaptive federated aggregation methods at the central server to resist poisoning attacks. Le et al. \cite{10184496} presented an adaptive privacy-preserving federated learning scheme, which implements an adaptive differential privacy strategy. Zhang et al. \cite{10136231} proposed an aggregation protocol incorporating the Mask technique to design a privacy-preserving scheme. Both \cite{10184496} and \cite{10136231} make use of the Euclidean distance to identify malicious updates, thereby effectively resisting poisoning attacks. 

Besides the threat of poisoning attacks, federated learning is also vulnerable to attacks by Byzantine nodes \cite{9468910,10070815,miao2022privacy,9615370,10293230,10098864,ruckel2022fairness,9761745}. In \cite{9468910}, Zhou et al. proposed a secure and efficient aggregation for Byzantine-robust federated learning to avoid Byzantine failures in distributed systems. Tao et al. \cite{10070815} developed  Byzantine-resilient federated learning at edge that realizes an order-optimal statistical error rate in the presence of Byzantine devices.  Miao et al. \cite{miao2022privacy} presented a privacy-preserving byzantine-robust federated learning to mitigate the impact of central servers and malicious clients. The scheme used cosine similarity to determine the malicious gradients uploaded by malicious clients. Wen et al. \cite{10293230} designed a Byzantine-resilient online federated learning algorithm to resist Byzantine attacks, which shows that their model's performance approaches an offline optimal classification model. Gouissem et al. \cite{10098864} proposed a low-complexity, decentralized Byzantine resilient training mechanism, which identifies and isolates hostile nodes instead of mitigating their impact on the global model. Xu et al. \cite{9761745} developed a blockchain-enabled secure and privacy-preserving decentralized learning system to resist malicious attacks from Byzantine nodes. 

A common drawback of the schemes mentioned above is that they all ignore the fundamental challenge of non-IID data, whose presence can make them inapplicable in real-world scenarios.

\subsection{Summary}
Different from the aforementioned studies, our approach (FedRFQ) introduces a secure prototype-based federated learning approach that ensures data security and improves model accuracy. Concretely, we use the SoftPool technology to solve the problems of prototype redundancy and prototype failure, thereby enhancing the ability of FedRFQ to handle non-IID data among clients. Additionally, we design a BFT detectable aggregation algorithm to filter out the low-quality prototypes, effectively resisting the poisoning attacks and safeguarding against server malfunctions. 

\section{Preliminaries and Models}\label{PRELIMINARIES}
In this section, we first introduce the preliminary knowledge that facilitates the design of FedRFQ, then present a clear outline of our models and design goals. 

\subsection{Preliminaries}
\textbf{Federated learning} is a framework for distributed deep learning with privacy. It enables a collective of clients to collaboratively train a global model based on their local data instead of transmitting the local data to a central server. During each round of the global model training, the central server distributes the latest global model to each client. Subsequently, each client proceeds to update its local model using the received global model, performs local model training, and fine-tunes its local model by gradient or weight information. Finally, the client submits its updated local model to the central server for aggregation, which ensures the privacy of all clients' data while improving the accuracy of the global model.

In a federated learning system, the collective set of clients is denoted as $K$, with the dataset at the $k$-th client represented by $\emph{D}_k = \{(x_{k,i},y_{k,i})_{i = 1}^{|\emph{D}_k|}\}$, where $(x_{k,i},y_{k,i})$ denotes a sample of dataset $\emph{D}_k$, with $x_{k,i}$ being the feature of the $i$-th sample at the $k$-th client, and $y_{k,i}$ the corresponding label. Each client independently conducts training on its local model, typically a deep neural network. When the training is complete, clients proceed to transmit their respective local models to a central server. Subsequently, the server employs an aggregation algorithm, such as FedAVG \cite{mcmahan2017communication}, to update the global weight tensor $\mathbf{\omega}$. The aggregation algorithm performs the following operation (\ref{pkwk}) to aggregate the local models:
\begin{equation}
\omega^{t+1} \longleftarrow \frac{1}{K} \sum^K_{k = 1}  \omega_{k}^t,
\label{pkwk}
\end{equation}
where $\omega_{k}^t$ denotes the weight of the local model of the $k$-th client in the $t$-th round. The global loss function is defined as $\mathcal{L}(f^+\left(\omega^t ; x\right),y) = \frac{1}{K} \sum^K_{k = 1}  \bm{\ell}_k(f^+\left(\omega^t ; x_{k}\right),y_{k})$, where $\bm{\ell}_k(\cdot)$ is the local loss function of the $k$-th client and $f^+(\cdot)$ denotes the local model. Generally,
\begin{equation}
\bm{\ell}_k(f^+\left(\omega^t ; x_{k}\right),y_k) \triangleq \frac{1}{ |\emph{D}_k|} \sum_{\left(x_{k,i},y_{k,i}\right) \in \emph{D}_k}^{|\emph{D}_k|} \ell_d(f^+\left(\omega^t ; x_{k,i}\right),y_{k,i}),
\end{equation}
where $\ell_d(\cdot)$ represents the loss function pertaining to the given sample $\left(x_{k,i},y_{k,i}\right)$. In federated learning, each client performs local training with the objective of minimizing its specific local loss function $\bm{\ell}_k(\cdot)$. The entire system is trained to minimize the global loss function $\mathcal{L}(\cdot)$: 
\begin{equation}
\begin{aligned}
 &\arg \min _{\omega^t}\left\{\mathcal{L}(f^+\left(\omega^t ; x\right),y)\right\} \\ &= \arg \min _{\omega^t}\left\{ \frac{1}{K} \sum^K_{k = 1}  \bm{\ell}_k(f^+(\omega^t ; x_{k}),y_k)\right\}.
 \end{aligned}
\end{equation}

However, in a real federated learning environment, each client collects data according to its own personal attributes, resulting in varied data classes and distinct data distributions among the clients. This diversity poses a challenge to heterogeneity in the federated learning setting. Generally, in statistical heterogeneity, prototype-based federated learning needs to recognize all classes. However, each client can recognize only a few classes. Moreover, classes overlap among the clients because different clients may have the same class. Therefore, to improve the tolerance of heterogeneity, we introduce a secure prototype-based federated learning approach on non-IID data in this paper. 

\textbf{Prototype} is the core part of the local training process in prototype-based federated learning. The local model consists of a representation layer and a decision layer, with the former being used to transform the raw data into high-level semantic information, which is known as a prototype. Let $f_k\left(\phi_k\right)$ denote the function of the $k$-th client's representation layer, where $\phi_k$ is the weight of the representation layer. We define $C_{k,i}$ = $f_k\left(\phi_k ; x_{k,i}\right)$ as the prototype of $x_{k,i}$. The decision layer can make predictions for the class of a given sample, and this layer consists of a fully connected layer and a softmax layer in FedRFQ. More specifically, the prediction for $x_{k,i}$ can be generated by the function $g_k\left(v_k;C_{k,i}\right)$ within the decision layer, where $v_k$ is the weight of the decision layer. We use $\omega_k$ to represent $(\phi_k,v_k)$. Consequently, the local model can be succinctly expressed as $f^+_k\left(\omega_k\right)=g_k\left( v_k\right) \circ f_k\left(\phi_k\right)$.

\subsection{Models and Design Goals}\label{model}
\begin{figure}[!htb]
	\centering
	\includegraphics[scale=0.46]{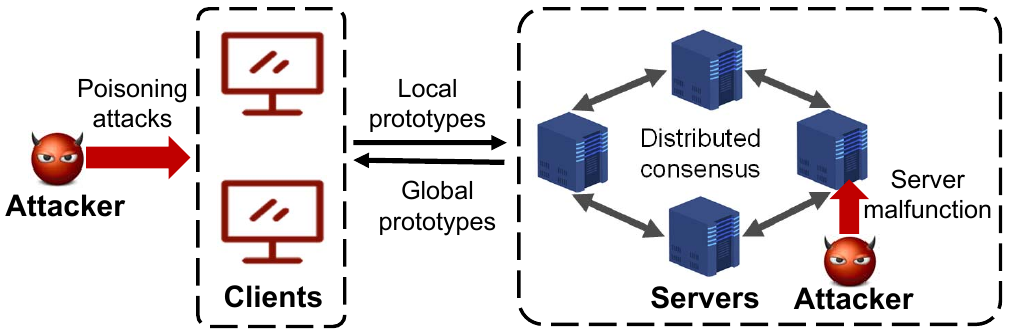}
	\caption{The model of FedRFQ.}
	\label{modelFedRFQ}
\end{figure}

\textbf{Network and Adversary Models.} The network model utilized in FedRFQ consists of two entities: clients and servers, as depicted in Fig.~\ref{modelFedRFQ}. Initially, each client trains its local prototypes using its own personal data. These prototypes are then transmitted to the servers, which play a critical role in verifying and aggregating the local prototypes. The servers subsequently compute and reach a consensus on the global prototype. The resulting global prototype is disseminated to the clients for model updating. Through multiple iterations, clients can converge on a final global model.

In our FedRFQ framework, the vulnerability of clients to poisoning attacks and the susceptibility of servers to malfunctions are also depicted in Fig.~\ref{modelFedRFQ}. We assume that dishonest clients masquerade as honest ones and execute poisoning attacks with the objective of generating erroneous prototypes. These attacks follow an untargeted strategy. However, to avoid being detected, malicious clients initiate poisoning attacks without excessively modifying the values of their prototypes, thereby further preventing the exposure of their own identity. The purpose of uploading these falsified prototypes is to disrupt the training process. Additionally, server malfunctions can result in the loss of locally stored prototypes. Moreover, malicious servers possess the capability to manipulate global prototypes, thereby disrupting the federated learning training process.

Nevertheless, we assume that proportion of malicious servers cannot exceed $1/3$ of the total. This assumption is critical for ensuring the robustness and reliability of FedRFQ. If malicious servers exceed $1/3$ of the total, they may send inconsistent information to the honest servers, thereby preventing the system from reaching a consensus.

\textbf{Design Goals.} FedRFQ, as a secure prototype-based federated learning on non-IID data, should be designed to achieve the following goals:
\begin{itemize}
\item Efficiency: FedRFQ aims to reduce the volume of parameters transmitted by the clients, thereby improving the overall communication efficiency and reducing prototype redundancy. 
\item Security: The core objective of FedRFQ is to resist poisoning attacks and server malfunctions while ensuring the robustness of FedRFQ. 
\item Model Accuracy: Confronting our adversary model and non-IID environments, FedRFQ still needs to enhance the accuracy across different datasets.
\end{itemize}

\section{FedRFQ}\label{FedProPool}
In this section, we present an overview on FedRFQ, followed by a comprehensive delineation on its construction. 

\subsection{Overview}
In FedRFQ, we integrate the SoftPool mechanism into prototype-based federated learning to address the challenges of prototype redundancy and prototype failure, thereby improving FedRFQ's efficiency in handling non-IID data. Furthermore, to enhance the security of FedRFQ, we introduce a BFT detectable aggregation algorithm, denoted as BFT-detect, which offers protection against poisoning attacks on clients and server malfunctions. 
\begin{figure}[!htb]
	\centering 
	\includegraphics[scale=0.4]{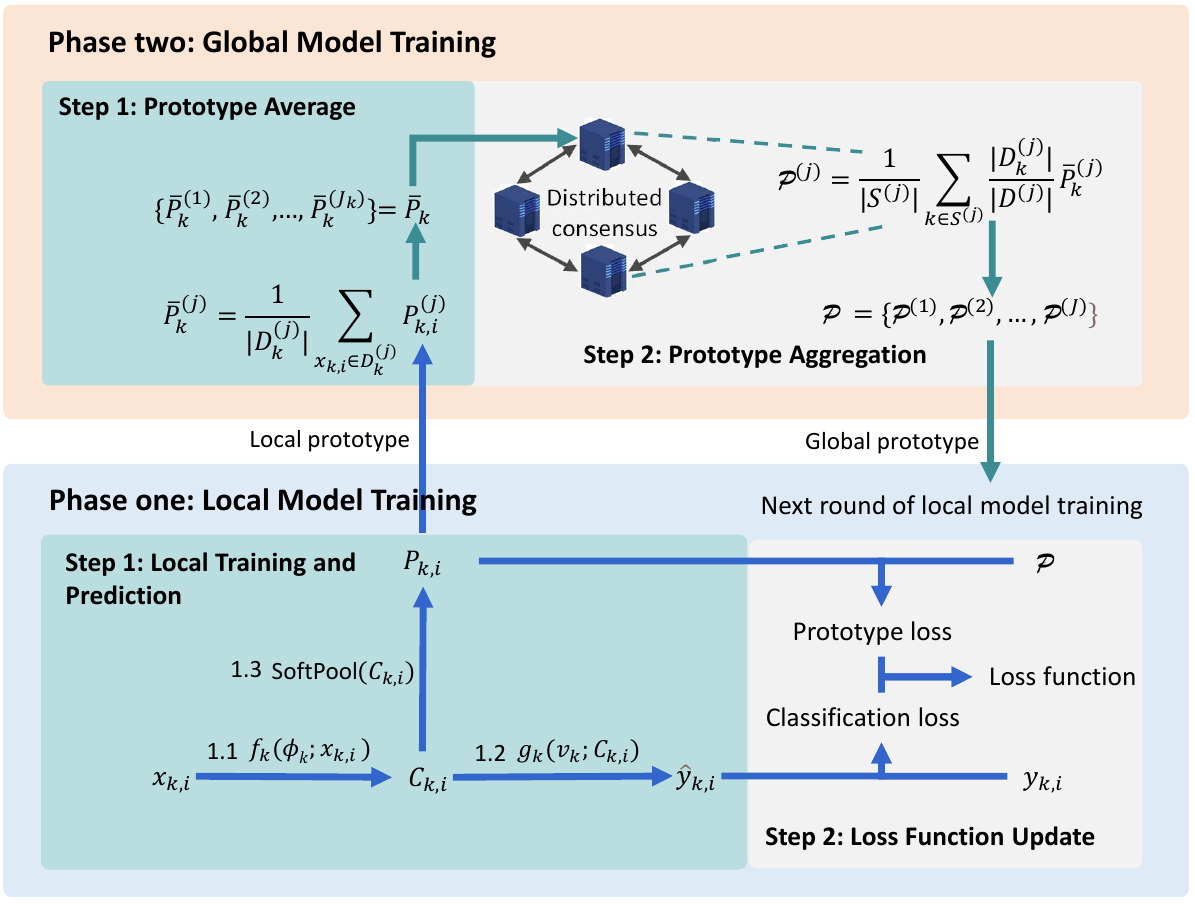}
	\caption{The one-round execution of FedRFQ.}
	\label{pool}
\end{figure}

We utilize a prototype to denote the feature representation of a sample in FedRFQ. For samples belonging to the same class, the average of their prototypes denotes the feature representation of that class. The training of FedRFQ includes multiple rounds. We take a one-round execution of the $k$-th client to illustrate the key steps of FedRFQ, which are shown in Fig.~\ref{pool}. Each round contains two phases, performing the tasks of local model training and global model training.

\textbf{Phase one}: \textit{Local Model Training}. Repeat the following two steps for $E$ iterations.
\begin{itemize} 
    \item \textbf{Step 1}: \textit{Local Training and Prediction.} Client $k$ randomly selects a subset of the local data, $\emph{D}_{k,e} \subset \emph{D}_{k}$, and performs the following three sub-steps for each $x_{k,i}$ in $\emph{D}_{k,e}$:
    
    \textbf{1.1}: \textit{Prototype Preparation (Representation Layer)}.  The $k$-th client obtains the prototype of $x_{k,i}$, i.e., $C_{k,i}$ = $f_k\left(\phi_k ; x_{k,i}\right)$, in the representation layer $f_k\left(\phi_k\right)$;

    \textbf{1.2}: \textit{Prediction (Decision Layer)}. The $k$-th client inputs the prototype $C_{k,i}$ in the decision layer to get the prediction $\hat{y}_{k,i}$  for $x_{k,i}$, i.e., $\hat{y}_{k,i}=g_k\left(v_k;C_{k,i}\right)$;

    \textbf{1.3}: \textit{SoftPool Embedding (Representation Layer)}. The $k$-th client pools prototype $C_{k,i}$ to obtain the pooled prototype $P_{k,i}$ by utilizing the SoftPool method. 

    \item \textbf{Step 2} \textit{Loss Function Update}. After obtaining $P_{k,i}$, the $k$-th client updates its local loss function (see Section \ref{SoftPool42} for details), considering both the classification loss incurred by $\hat{y}_{k,i}$ and the prototype loss caused by the difference of $P_{k,i}$ and the global prototype from the previous round,  thereby updating $\phi_k$ and $v_k$ to get a new $\omega_k$.  
\end{itemize}

Note that the number of iterations $E$ to repeat Steps 1 and 2 in Phase one depends on the dataset size of the client and specific machine learning tasks. $E$ should be sufficiently large to ensure an acceptable model accuracy. Also note that the updates of local loss functions at the clients effectively update the global loss function, which can be regarded as the global objective.

    \textbf{Phase two}: \textit{Global Model Training}. 

    \begin{itemize}
    
    \item \textbf{Step 1}: \textit{Prototype Average}. After Phase one, client $k$ calculates the average of the pooled prototypes. Concretely, each sample $x_{k,i}$ in client $k$ belongs to a class $y_{k,i}$. Then for each class, client $k$ calculates the average of the pooled prototypes of the samples belonging to that class, and finally obtains the set of averaged prototypes for all classes, denoted as $\bar{P}_k$. Client $k$ uploads the set $\bar{P}_k$ to the servers.
 
    \item \textbf{Step 2}: \textit{Prototype Aggregation}. Upon receiving the sets of averaged prototypes from all clients, servers execute the BFT-detect mechanism to filter out low-quality prototypes and the leader in BFT-detect aggregates the received prototypes to obtain the global prototype set $\mathcal{P}$. Then, the servers distribute the global prototype $\mathcal{P}$ to each client, which will be used for local model training in the next round.

\end{itemize}

\subsection{Key Steps of FedRFQ}\label{SoftPool42}
In this subsection, we detail the key steps of FedRFQ, focusing on the operations of one round. Note that we describe the loss functions at the end for better elaborations.
    
    \textit{1) Local Training and Prediction:} During this process, the representation layer $f_k(\phi_k)$ extracts the crucial features of $x_{k,i}$ based on $\phi_k$ to get the prototype $C_{k,i}$, i.e., $C_{k,i} = f_k\left(\phi_k; x_{k,i}\right)$; then the decision layer apply the function $g_k\left(v_k;C_{k,i}\right)$ to obtain the prediction $\hat{y}_{k,i}$ for $x_{k,i}$, i.e., $\hat{y}_{k,i}=g_k\left(v_k;C_{k,i}\right)$.

\begin{figure*}
	\centering 
	\includegraphics[scale=0.55]{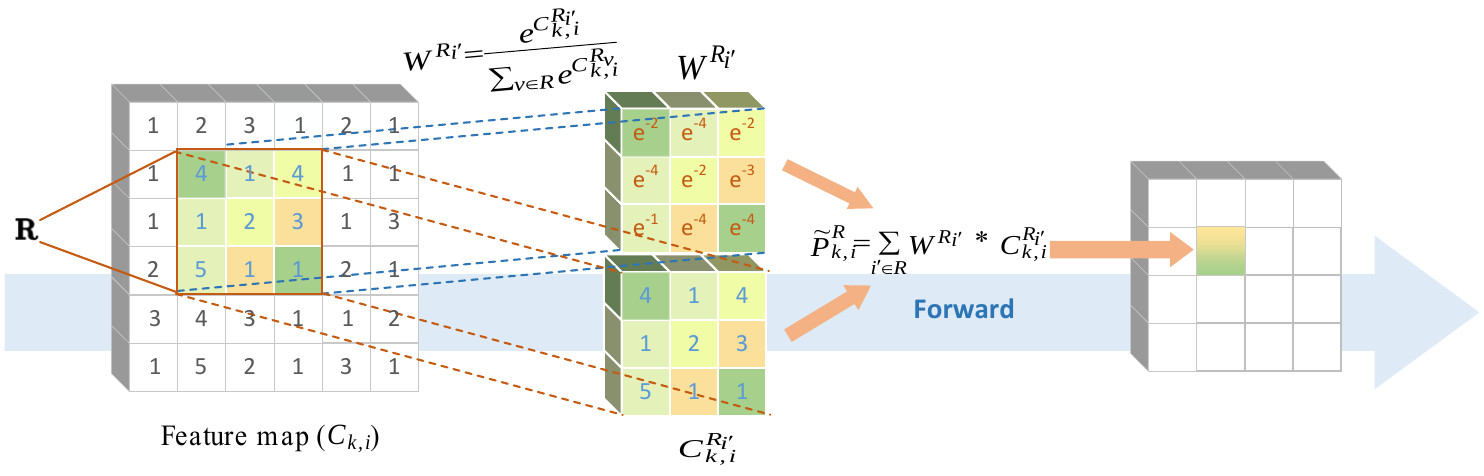}
	\caption{The pooling operation with SoftPool}
	\label{softpool}
\end{figure*}

    To solve the problems of prototype redundancy and prototype failure, we embed the SoftPool pooling method in \cite{stergiou2021refining} into the training process. Specifically, the prototype $C_{k,i}$ is regarded as a 2D feature map, as shown in Fig. \ref{softpool}. We define a pooling filter with a size of $\hat{k} \times\hat{k}$ to pool the whole map. Let $\mathbf{R}$ be the kernel region covered by the pooling filter, we have $|\mathbf{R}| = \hat{k}^2$.   The output of the pooling operation is $\tilde{P}^\mathbf{R}$. The pooling stages are presented as follows and illustrated in Fig. \ref{softpool}. 

\textbf{Stage 1}: In the kernel region $\mathbf{R}$, SoftPool adopts a smooth maximum approximation. We denote $C_{k,i}^{\mathbf{R}_{i^{\prime}}}$ as the $i^\prime$-th feature value in the kernel region $\mathbf{R}$. The weight $W^{\mathbf{R}_{i^{\prime}}}$ of $C_{k,i}^{\mathbf{R}_{i^{\prime}}}$ is obtained by Eq. (\ref{e6}), which represents the ratio of the natural exponent of that feature value with the sum of all natural exponents of the feature values in the kernel region $\mathbf{R}$.
\begin{equation}
W^{\mathbf{R}_{i^{\prime}}}=\frac{e^{C_{k,i}^{\mathbf{R}_{i^{\prime}}}}}{\sum_{v \in \mathbf{R}} e^{C_{k,i}^{\mathbf{R}_v}}}.
	\label{e6}
\end{equation}

\textbf{Stage 2}: To capture crucial patterns of the feature values, the output $\tilde{P}_{k,i}^\mathbf{R}$ of the SoftPool operation is produced by a standard summation of all weighted feature values in the kernel region $\mathbf{R}$ by Eq. (\ref{ee12}).
\begin{equation}
	\tilde{P}_{k,i}^\mathbf{R} =\sum_{i^{\prime} \in \mathbf{R}} W^{\mathbf{R}_{i^{\prime}}} * C_{k,i}^{\mathbf{R}_{i^{\prime}}}.\label{ee12}
\end{equation}

After the current kernel region $\mathbf{R}$ completes the pooling operation, the filter moves to the next kernel region $\mathbf{R_1}$ for the pooling operation. Stage $1$ and Stage $2$ are repeated until all regions in prototype $C_{k,i}$ have been pooled.
The final output values $\{\tilde{P}_{k,i}^\mathbf{R},\tilde{P}_{k,i}^\mathbf{R_{1}},...\}$ are combined according to their pooling positions to obtain a pooled prototype $P_{k,i}$. We simplify the above operations into Eq. (\ref{ee1}).
\begin{equation}
	P_{k,i} = \mathtt{SoftPool}(C_{k,i}).
	\label{ee1}
\end{equation}

\textit{2) Prototype Average:} The $k$-th client obtains $P_{k,i}$ for each sample $x_{k,i}$ in the dataset $D_k$. Then it calculates averaged prototypes by class. For the $j$-th class, we can obtain the $\bar{P}_k^{(j)}$ through Eq. (\ref{pingjun}), where $ j \in \{1,2,..., J\}$. 
\begin{equation}
	\bar{P}_k^{(j)}=\frac{1}{\left|\emph{D}_{k}^{(j)}\right|}\sum_{x_{k,i} \in \emph{D}^{(j)}_{k}}P_{k,i}^{(j)},
	\label{pingjun}
\end{equation}
where $\emph{D}_{k}^{(j)}\in D_k$ is the set of samples belonging to the $j$-th class for client $k$. Then client $k$ obtains the set of averaged prototypes for all classes $\bar{P}_k =\{ \bar{P}_k^{(1)},\bar{P}_k^{(2)},...,\bar{P}_k^{(J_k)}\}$,
and uploads them to the servers. 

    \textit{3) Prototype Aggregation:} For simplicity, we circumvent the BFT-detect (details in Section~\ref{Consensus}) here but illustrate a generic prototype aggregation method in SoftPool-based learning. After receiving $\bar{P}_k$ from all clients, the server aggregates them and obtains the global prototype for each class $j$ by the following operation (\ref{eq9}),
\begin{equation}
	\mathcal{P}^{(j)}=\frac{1}{\left|S^{(j)}\right|} \sum_{k \in S^{(j)}} \frac{\left|\emph{D}_{k}^{(j)}\right|}{|\emph{D}^{(j)}|} \bar{P}_k^{(j)},
	\label{eq9}
\end{equation}
where $S^{(j)}$ denotes the set of clients that have the $j$-th class, and $\emph{D}^{(j)}$ denotes the set of samples belonging to the $j$-th class across all clients. Denote by $\mathcal{P} =\{ \mathcal{P}^{(1)},\mathcal{P}^{(2)},..., \mathcal{P}^{(J)}\}$ the set of global prototypes for all classes.  $\mathcal{P}$ is then sent to each client, which is used for local model training in the next round.

    \textit{4) Loss Function Update:} To improve the effectiveness of the local model, one needs to minimize the local loss function $\bm{\ell}_k$. As the goal of this step is to minimize the classification loss for each client while ensuring the alignment of the local prototype with the global one, the loss function consists of two components: the classification loss and the prototype loss between the local prototype $P_{k,i}$ in the current round and the global prototype $\mathcal{P}$ from the previous round. Accordingly we have    
\begin{equation}
\begin{aligned}
	& \bm{\ell}_k\left(f^+\left(\omega_k; x_{k}\right),y_k\right) \\ &= \frac{1}{ |\emph{D}_{k,e}|} \sum_{\left(x_{k,i},y_{k,i}\right) \in \emph{D}_{k,e}}^{|\emph{D}_{k,e}|} \ell_d(f^+\left(\omega_k; x_{k,i}\right),y_{k,i}) \\ &+\lambda \cdot \sum^J_{j} \left\|P_{k}^{(j)}-\mathcal{P}^{(j)}\right\|_2,
\end{aligned}
	\label{lof88}
\end{equation}
where
\begin{equation}
\left\|P_k^{(j)}-\mathcal{P}^{(j)}\right\|_2 = \frac{1}{ |\emph{D}_{k,e}^{(j)}|} \sum_{\left(x_{k,i},y_{k,i}\right) \in \emph{D}_{k,e}^{(j)}}^{|\emph{D}_{k,e}^{(j)}|}  \left\|P_{k,i}^{(j)}-\mathcal{P}^{(j)}\right\|_2.
	\label{lofaaaa}
\end{equation}

In Eq.(\ref{lof88}), $\lambda$ represents the weight of prototype loss with respect to the global prototype $\mathcal{P}^{(j)}$. $\left\|P_k^{(j)}-\mathcal{P}^{(j)}\right\|_2$ obtained by Eq. (\ref{lofaaaa}) denotes the $L_2$ distance between local prototype $P_{k}^{(j)}$ and global prototype $\mathcal{P}^{(j)}$. Then the $k$-th client calculates the gradient of the loss function at each iteration  by 
\begin{equation}
    g^{e}_{k} = \nabla \bm{\ell}_k(\omega_{k}^{e-1};\emph{D}_{k,e}),
    \label{gradient}
\end{equation}
where $e \in \{1,2,\cdots,E\}$, the symbol $\nabla$ represents the derivation operation. Then, the weight $\omega_{k}$ can be updated by Eq. (\ref{weight}).
\begin{equation}
\omega_{k}^{e}  = \omega_{k}^{e-1} - \eta g_{k}^{e}\label{weight}.
\end{equation} 

With the update of the local loss function at each client, the global loss function gets updated effectively to realize the objective of minimizing the overall classification loss and prototype loss of the entire system. It can be modeled mathematically as:
\begin{equation}
		\begin{aligned}
			\underset{\omega_1,\omega_2, \dots,\omega_K}{\arg \min } \{  
			&\sum_{k=1}^K \frac{1}{|\emph{D}|}  \sum_{\left(x_{k,i},y_{k,i}\right) \in \emph{D}_{k}}^{|\emph{D}_{k}|} \ell_d(f^+\left(\omega_k; x_{k,i}\right),y_{k,i}) \\
            & + \lambda \cdot \sum_{j=1}^{J} \sum_{k=1}^K \frac{\left|\emph{D}_{k}^{(j)}\right|}{|\emph{D}^{(j)}|}\left\|P_{k}^{(j)}-\mathcal{P}^{(j)}\right\|_2\},
   \label{gqf}
		\end{aligned}
	\end{equation}
where $|\emph{D}|$ denotes the number of samples on all clients,  $\sum_{k=1}^K \frac{1}{|\emph{D}|}  \sum_{\left(x_{k,i},y_{k,i}\right) \in \emph{D}_{k}}^{|\emph{D}_{k}|} \ell_d(f^+\left(\omega_k; x_{k,i}\right),y_{k,i})$ the sum of classification losses for all clients; and $\sum_{j=1}^{J} \sum_{k=1}^K \frac{\left|\emph{D}_{k}^{(j)}\right|}{|\emph{D}^{(j)}|} \left\|P_{k}^{(j)}-\mathcal{P}^{(j)}\right\|_2$ the sum of prototype losses for all clients. 

\subsection{BFT-detect}\label{Consensus}
To enhance the resistance against poisoning attacks and minimize the impact of server malfunctions, FedRFQ makes use of a BFT detectable aggregation algorithm called BFT-detect, which is designed to realize two objectives. First, it can detect and exclude low-quality prototypes, thereby preventing poisoning attacks; second, it utilizes a Practical Byzantine Fault-Tolerance (PBFT) algorithm to reach a consensus on the aggregation results, thereby protecting against server malfunctions. A detailed explanation on the algorithm is provided below.

\textit{1) Prototype Broadcasting:} When clients submit prototypes, the leader server (selected by random robin) creates a proposal that includes all the prototype information. This proposal is then shared with other servers to reach a consensus. If the leader makes a mistake, a view change mechanism is activated, which randomly selects a new leader for the next round of consensus. Because the number of Byzantine servers in the network doesn't exceed $N/3$, where $N$ is the total number of servers, switching up to $O(N)$ times can obviously lead to a correct leader being selected.

\textit{2) Prototype Quality Detection:} After each server receives the set $\bar{P}_k$ from each client, the leader server performs quality detection to filter out low-quality prototypes according to the following process:
\begin{itemize}
\item For each class, the leader calculates $\mathcal{P}^{(j)}$ by Eq. (\ref{eq9}) and obtains $\mathcal{P} =\{ \mathcal{P}^{(1)},\mathcal{P}^{(2)},..., \mathcal{P}^{(J)}\}$. 

\item Then, the leader calculates the discrepancy $d_k$ between  $\bar{P}_k$ and $\mathcal{P}$ by $L_2$ distance. The set of discrepancies for all clients is denoted as $L = \{d_1, d_2, \ldots, d_K\}$.
Note that if the prototype set $\bar{P}_k$ at the $k$-th client does not have an averaged prototype for the $j$-th class, the $j$-th class doesn't participate in the computation of discrepancy.

\item In BFT-detect, we set a security level $\psi$ for filtering low-quality prototypes. Specifically, we filter out all the prototypes of $\psi$ clients whose $L_2$ distances are the largest $\psi$ values in $L$. 
\end{itemize}

\textit{3) Global Prototype Calculation:} After the leader executes the prototype quality detection mechanism, the local prototypes that are retained are used to calculate the set of global prototypes for all classes. Subsequently, the leader proposes the global prototype set, which is then broadcast to other servers for proposal verification.

\textit{4) Data Verification:} When a leader presents a proposal of a filtered global prototype set, all servers verify it based on the local result of prototype quality detection and cast their ``prepare'' votes for the proposal. Each server affixes a digital signature to a vote before broadcasting it to other servers. If a server receives $2f+1$ ``prepare'' votes for the proposal from different servers, the server casts a ``commit'' vote for the proposal, where $f$ denotes the number of malicious servers. If a server receives $2f+1$ ``commit'' votes for the proposal from different servers, the proposal is confirmed.

\section{Analysis}\label{THEORETICAL}
We first analyze the upper bound of the local loss function of FedRFQ. This analysis serves as an instruction of FedRFQ's robustness in the face of poisoning attacks launched by malicious clients. Then, we evaluate a server's ability to avoid malfunctions.

\subsection{Upper Bound Analysis on the Loss Function with Poisoning Attacks}
Following the general studies \cite{wang2020tackling,li2020federated}, we make the following assumptions on the local loss function $\bm{\ell}_k$, and
add a new superscript $\tau$ to denote the index of iterations across all rounds. In the ($t+1$)-th round, $\tau$ and $e$ satisfy $\tau=tE+e$, where $t,e=0,1,\cdots,$ with $t$ denoting the index of rounds and $e$ the index of iterations within a round.

\begin{assumption}
	(Lipschitz Smooth) Each local loss function is $L_a$-Lipschitz smooth, which means that the local loss function is smooth and its gradient conforms to the $L_a$-Lipschitz continuity:
	\begin{equation}
		\begin{gathered}
			\left\|\nabla \bm{\ell}_k^{\tau_1}-\nabla \bm{\ell}_k^{\tau_2}\right\|_2 \leq L_a\left\|\omega_{k}^{\tau_1}-\omega_{k}^{\tau_2}\right\|_2, \\
			\forall \tau_1, \tau_2>0, k \in\{1,2, \ldots, K\}.
		\end{gathered}
     \label{Laequation}
	\end{equation}
 
  In Eq. (\ref{Laequation}), for each client, there exists a value $L_a$ to make it hold. In the $\tau_1$-th iteration, $\nabla \bm{\ell}_k^{\tau_1}$ denotes the first-order derivative of the loss function and $\omega_{k}^{\tau_1}$ is the model weight. Similarly, in the $\tau_2$-th iteration, $\nabla \bm{\ell}_k^{\tau_2}$ denotes the first-order derivative of the loss function and $\omega_{k}^{\tau_2}$ is the model weight.   
	Its equivalent condition is a quadratic upper bound,	
		\begin{equation}
			\begin{aligned}
		\bm{\ell}_k^{\tau_1}-\bm{\ell}_k^{\tau_2} \leq & \left\langle\nabla \bm{\ell}_k^{\tau_2},\left(\omega_{k}^{\tau_1}-\omega_{k}^{\tau_2}\right)\right\rangle \\
  & +\frac{L_a}{2}\left\|\omega_{k}^{\tau_1}-\omega_{k}^{\tau_2}\right\|_2^2, \\
				& \forall \tau_1, \tau_2>0, \quad k \in\{1,2, \ldots, K\},
			\end{aligned}
\nonumber
		\end{equation}
  where $\bm{\ell}_k^{\tau_1}$ denotes the loss function in the $\tau_1$-th iteration and $\bm{\ell}_k^{\tau_2}$ denotes the loss function in the $\tau_2$-th iteration.
	\label{assumption1}
\end{assumption}
\begin{assumption}
	(Unbiased Gradient and Bounded Variance) The unbiased estimator of the gradient for each client is a stochastic gradient  $g_{k}^{\tau}=\nabla \bm{\ell}_k\left(\omega_{k}^{\tau}; x_{k}\right)$. Assume that its expectation is
	\begin{equation}
		\mathbb{E}_{x_{k} \sim D_k}\left[g_{k}^{\tau}\right]=\nabla \bm{\ell}_k\left(\omega_{k}^{\tau}\right)=\nabla \bm{\ell}_k^{\tau}, \forall k \in\{1,2, \ldots, K\},
		\label{eq15}
	\end{equation}
   where $\nabla \bm{\ell}\left(\omega_{k}^{\tau}\right)$ represents the first-order derivative of the loss function $\bm{\ell}_k$ with the model weight $\omega_{k}^{\tau}$. The variance of expectation of stochastic gradient $g_{k}^{\tau}$ is bounded by $\sigma^2$. We have
	\begin{equation}
		\mathbb{E}\left[\left\|g_{k}^{\tau}-\nabla \bm{\ell}\left(\omega_{k}^{\tau}\right)\right\|_2^2\right] \leq \sigma^2, \forall k \in\{1,2, \ldots, K\}, \sigma^2 \geq 0.
		\label{ee3}
	\end{equation}
	\label{assumption2}
\end{assumption}

\begin{assumption}
	(Bounded Expectation of Stochastic Gradients) For each client $k \in \{1, 2, \ldots, K\}$, the expectation of the Euclidean norm of the stochastic gradient $g_{k}^{\tau}$ is bounded by $G$:
	\begin{equation}
	\mathbb{E}\left[\left\|g_{k}^{\tau}\right\|_2\right] \leq G.
  \nonumber
	\end{equation}
	\label{assumption3}
\end{assumption}
\begin{assumption}
	(Lipschitz Continuity). Each representation layer is $L_2$-Lipschitz continuous, 
	\begin{equation}
		\begin{gathered}
			\left\|f_k\left(\phi_{k}^{\tau_1}\right)-f_k\left(\phi_{k}^{\tau_2}\right)\right\|_2 \leq L_2\left\|\phi_{k}^{\tau_1}-\phi_{k}^{\tau_2}\right\|_2, \\
			\forall \tau_1, \tau_2>0, k \in\{1,2, \ldots, K\},
		\end{gathered}
   \label{L2equation}
	\end{equation}
 where $f_k\left(\phi_{k}^{\tau_1}\right)$ denotes the function of the representation layer and $\phi_{k}^{\tau_1}$ is its weight in the $\tau_1$-th iteration. In the $\tau_2$-th iteration, $f_k\left(\phi_{k}^{\tau_2}\right)$ denotes the function of the representation layer and $\phi_{k}^{\tau_2}$ is its weight.
	\label{assumption4}
\end{assumption}

Based on the aforementioned assumptions, Theorem \ref{theorem1} provides an upper bound of the local loss function between two rounds of communications. 

\begin{theorem}
	(The upper bound) For any client, after each round of communications in FedRFQ, $\mathbb{E}\left[\bm{\ell}_k^{((t+1)E)^\prime}\right]$ is upper bounded by
 \begin{footnotesize}
	\begin{equation}
		\begin{aligned}
			\mathbb{E}\left[\bm{\ell}_k^{((t+1)E)^\prime}\right] & \leq  \bm{\ell}_k^{(tE)^\prime} \\ &-\left(\eta-\frac{L_a \eta^2}{2}\right) (\sum_{e=1}^{E-1}\left\|\nabla \bm{\ell}_k^{tE+e}\right\|_2^2  + \left\|\nabla \bm{\ell}_k^{(tE)\prime}\right\|_2^2)\\
			& +\frac{L_a E \eta^2}{2} \sigma^2+\lambda L_2 L^{\prime}_2 \eta E G+ 2\varepsilon,
		\end{aligned}
	\end{equation}
 \end{footnotesize}
where $\bm{\ell}_k^{(tE)^\prime}$ denotes the local loss function after prototype aggregation in the $t$-th round. Similarly, $\bm{\ell}_k^{((t+1)E)^\prime}$ denotes the local loss function when competing the prototype aggregation in the ($t+1$)-th round. $\lambda$ represents the weight of prototype loss, $\eta$ represents the learning rate of a client and $2\varepsilon$ denotes the gap caused by embedding the quality detection mechanism in the prototype aggregation. And there exists a value $L_2^{\prime}$, which can ensure that $\textit{SoftPool}$ satisfies $L_2^{\prime}$-Lipschitz continuity. Theorem \ref{theorem1} establishes an upper bound of the local loss function after each round for any client, and its proof is provided in Appendix \ref{theorem111}. In addition, SoftPool also satisfies $L_2^{\prime}$-Lipschitz continuity, and the detailed proof of this property is presented in Appendix \ref{Proof111}. Thus, we can conclude that in the presence of poisoning attacks, FedRFQ ensures that the loss function has an upper bound, demonstrating the robustness of FedRFQ. Further, according to Corollary 1 in the study \cite{tan2022fedproto},  by calculating appropriate values for $\eta$, $\lambda$ and $\varepsilon$, we can ensure the convergence of the local loss function. 
	\label{theorem1}
\end{theorem}

\subsection{Resistance against Server Malfunction}
\begin{figure}[!htb]
	\centering 
	\includegraphics[width=3.5in,height=1.9in]{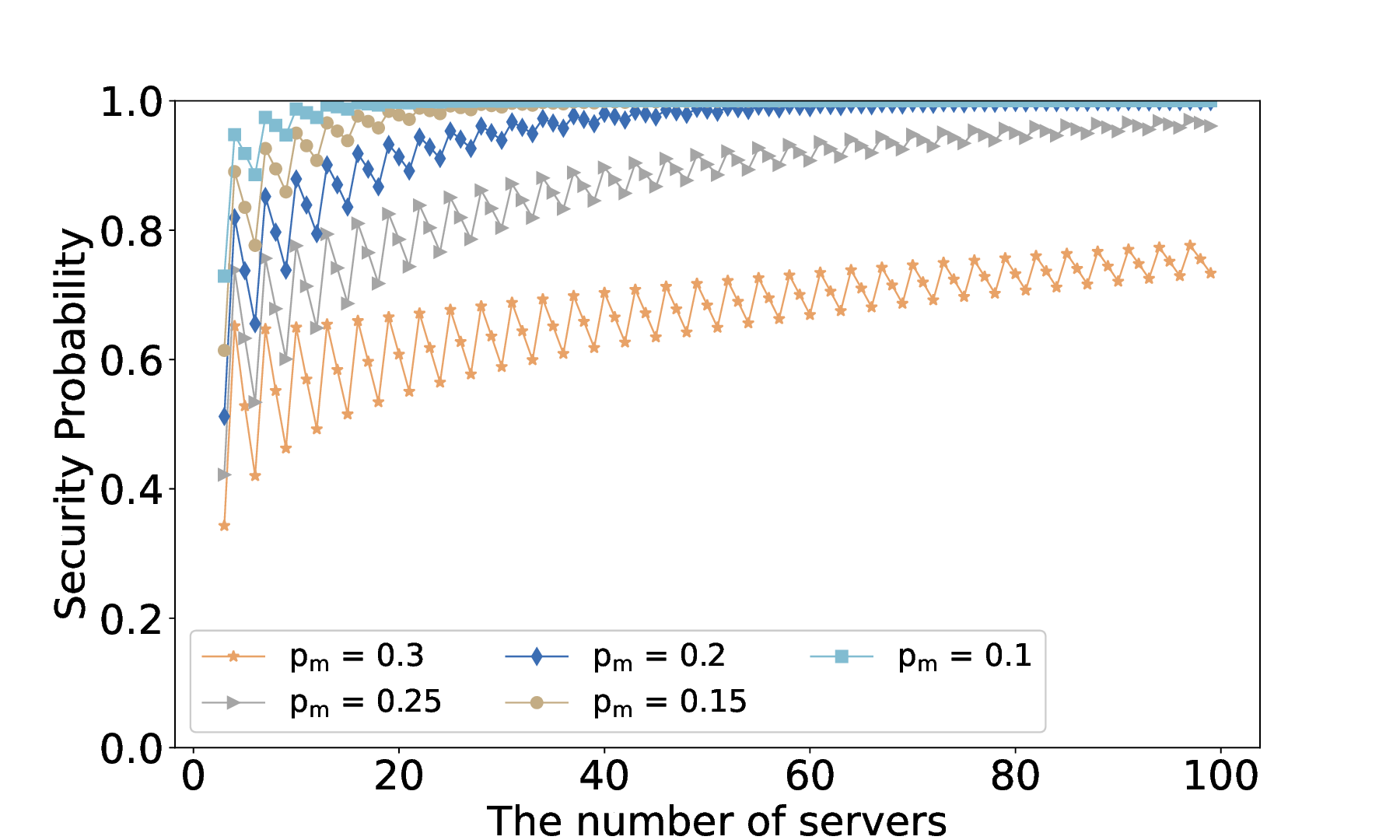}
	\caption{The security probability with respect to $N$, the total number of servers, under different probability of a server being malicious.}
	\label{securepro}
\end{figure}

The security of BFT-detect in FedRFQ depends on whether $f \leq \lfloor\frac{N-1}{3}\rfloor$ is satisfied, which means that FedRFQ can resist server malfunctions if $f \leq \lfloor\frac{N-1}{3}\rfloor$ holds. Here, $N$ is the total number of servers and $f$ is the number of malicious servers. However, sometimes $f$ can be greater than $\lfloor\frac{N-1}{3}\rfloor$, which exceeds the optimal resiliency. To measure the effectiveness of FedRFQ in mitigating such bad cases, we utilize a metric known as security probability, which represents the probability on which the BFT-detect algorithm meets the condition $f \leq \lfloor\frac{N-1}{3}\rfloor$. We can calculate the security probability by $\sum_{i=0}^{f_{max}}\left(\begin{array}{l}N\\i\end{array}\right) \mathsf{P}_\mathsf{m}^i(1-\mathsf{P_m})^{N-i}$, where $\mathsf{P_m}$ refers to the probability of a server being malicious, and $f_{max} = \lfloor\frac{N-1}{3}\rfloor$ for a fixed $N$. 

Fig.~\ref{securepro} displays the simulation results of our investigation on security probability. In this experiment, we set $\mathsf{P_m}$ with a value ranging from 0.1 to 0.3, and observe that the security probability of the BFT-detect consensus process decreases as the probability of a server becoming malicious increases. However, regardless of how the probability $\mathsf{P_m}$ changes, an increase in the total number of servers results in a higher security probability. This is because a larger number of servers leads to a more secure consensus process and more honest servers participate in proposal verification. 
When $\mathsf{P_m}<0.1$ and the number of servers exceeds $20$, the security probability exceeds 0.99. Thus, one can conclude that the BFT-detect consensus algorithm is reliable and secure, provided that the value of $\mathsf{P_m}$ is relatively small and the framework has a sufficient number of servers participating in consensus.

\begin{table*}[!htb]
\centering
\caption{Comparison of federated learning methods on the MNIST dataset under non-IID settings. The experimental results for FedRFQ come from the following parameter settings $\eta=0.01, \lambda=1, \zeta=0, \psi=0$ since these values can achieve better results through our continuous test. The best results are shown in bold. The results indicate that FedRFQ achieves higher accuracy with fewer communication parameters.}
\renewcommand\arraystretch{1.03}
\begin{tabular}{|c|c|c|c|c|c|c|c|}
	\hline \multirow{2}{*}{ Dataset } & \multirow{2}{*}{ Method } & \multirow{2}{*}{$\begin{array}{c}\mathsf{std} \\
			\text { of } \mathsf{avg}\end{array}$} & \multicolumn{3}{|c|}{ Test Accuracy } & \multirow{2}{*}{$\begin{array}{c}\text {Communication } \\
			\text{ Rounds }\end{array}$} & \multirow{2}{*}{$\begin{array}{c}\text{ Communication }\\ \text{ Params }\left(\times 10^3\right)\end{array}$} \\
	\cline{4-6} & & & $\mathsf{avg}=3$ & $\mathsf{avg}=4$ & $\mathsf{avg}=5$ & & \\
	\hline \multirow{14}{*}{MNIST} & Local & $\begin{array}{l}2 \\
		3\end{array}$ & $\begin{array}{l}94.05 \pm 2.93 \\
		93.44 \pm 3.57\end{array}$ & $\begin{array}{l}93.35 \pm 3.26 \\
		94.24 \pm 2.49\end{array}$ & $\begin{array}{l}92.92 \pm 3.17 \\
		93.97 \pm 2.97\end{array}$ & 0 & 0 \\
	\cline{2-8} & FeSEM \cite{xie2021multi} & $\begin{array}{l}2 \\
		3\end{array}$ & $\begin{array}{l}95.26 \pm 3.48 \\
		96.40 \pm 3.35\end{array}$ & $\begin{array}{l}97.06 \pm 2.72 \\
		95.82 \pm 3.94\end{array}$ & $\begin{array}{l}96.31 \pm 2.41 \\
		95.98 \pm 2.46\end{array}$ & 150 & 430 \\
	\cline{2-8} & FedProx \cite{li2020federated} & $\begin{array}{l}2 \\
		3\end{array}$ & $\begin{array}{l}96.26 \pm 2.89 \\
		96.65 \pm 3.28\end{array}$ & $\begin{array}{l}96.40 \pm 3.33 \\
		95.25 \pm 3.73\end{array}$ & $\begin{array}{l}95.65 \pm 3.38 \\
		95.34 \pm 2.85\end{array}$ & 110 & 430 \\
	\cline{2-8} & FedPer \cite{arivazhagan2019federated} & $\begin{array}{l}2 \\
		3\end{array}$ & $\begin{array}{l}95.57 \pm 2.96 \\
		96.57 \pm 2.65\end{array}$ & $\begin{array}{l}96.44 \pm 2.62 \\
		95.93 \pm 2.76\end{array}$ & $\begin{array}{l}95.55 \pm 3.13 \\
		96.07 \pm 2.80\end{array}$ & 100 & 106 \\
	\cline{2-8} & FedAvg \cite{mcmahan2017communication} & $\begin{array}{l}2 \\
		3\end{array}$ & $\begin{array}{l}91.40 \pm 6.48 \\
		94.57 \pm 4.91\end{array}$ & $\begin{array}{l}94.32 \pm 4.89 \\
		91.99 \pm 6.89\end{array}$ & $\begin{array}{l}93.22 \pm 4.39 \\
		92.19 \pm 3.97\end{array}$ & 150 & 430 \\
	\cline{2-8} & FedRep \cite{collins2021exploiting} & $\begin{array}{l}2 \\
		3\end{array}$ & $\begin{array}{l}94.96 \pm 2.78 \\
		95.01 \pm 3.92\end{array}$ & $\begin{array}{l}95.18 \pm 3.80 \\
		95.55 \pm 2.79\end{array}$ & $\begin{array}{l}94.94 \pm 2.81 \\
		95.38 \pm 2.97\end{array}$ & 100 & 110 \\
	\cline{2-8} & FedProto \cite{tan2022fedproto} & $\begin{array}{l}2 \\
		3\end{array}$ & $\begin{array}{l}97.13 \pm 0.30 \\
		96.71 \pm 0.43\end{array}$ & $\begin{array}{l}96.80 \pm 0.41 \\
		96.87 \pm 0.28\end{array}$ & $\begin{array}{l}96.70 \pm 0.29 \\
		96.47 \pm 0.23\end{array}$ & 100 & 4 \\
	\cline{2-8} & \textbf{FedRFQ} & $\begin{array}{l}2 \\
		3\end{array}$ & $\begin{array}{l}\textbf{98.41} \pm \textbf{0.25} \\
		\textbf{97.90} \pm \textbf{0.31}\end{array}$ & $\begin{array}{l}\textbf{97.67} \pm \textbf{0.32} \\
		\textbf{97.93} \pm \textbf{0.27}\end{array}$ & $\begin{array}{l}\textbf{97.40} \pm \textbf{0.28} \\
		\textbf{97.17} \pm \textbf{0.25}\end{array}$ & 100 & \textbf{1.92} \\
	\hline
\end{tabular}
\label{table1}
\end{table*}

\begin{table*}[!htb]
\centering
\caption{Comparison of federated learning methods on the FEMNIST dataset under non-IID settings. The experimental results for FedRFQ come from the following parameter settings  $\eta=0.01, \lambda=1, \zeta=0, \psi=0$ since these values can achieve better results through our continuous test. The best results are shown in bold. The results indicate that FedRFQ achieves higher accuracy with fewer communication parameters.}
\renewcommand\arraystretch{1.03}
	\begin{tabular}{|c|c|c|c|c|c|c|c|}
		\hline \multirow{2}{*}{ Dataset } & \multirow{2}{*}{ Method } & \multirow{2}{*}{$\begin{array}{c}\mathsf{std} \\
				\text { of } \mathsf{avg}\end{array}$} & \multicolumn{3}{|c|}{ Test Accuracy } & \multirow{2}{*}{$\begin{array}{c}\text { Communication } \\
				\text { Rounds }\end{array}$} & \multirow{2}{*}{$\begin{array}{c}\text { Communication } \\
				\text { Params }\left(\times 10^3\right)\end{array}$} \\
		\cline{4-6} & & & $\mathsf{avg}=3$ & $\mathsf{avg}=4$ & $\mathsf{avg}=5$ & & \\
		\hline \multirow{14}{*}{FEMNIST} & Local & $\begin{array}{l}1 \\
			2\end{array}$ & $\begin{array}{c}92.50 \pm 10.42 \\
			92.11 \pm 6.02\end{array}$ & $\begin{array}{l}91.16 \pm 5.64 \\
			90.34 \pm 6.42\end{array}$ & $\begin{array}{l}87.91 \pm 8.44 \\
			89.70 \pm 6.33\end{array}$ & 0 & 0 \\
		\cline{2-8} & FeSEM \cite{xie2021multi} & $\begin{array}{l}1 \\
			2\end{array}$ & $\begin{array}{l}93.39 \pm 6.75 \\
			94.19 \pm 4.90\end{array}$ & $\begin{array}{l}91.06 \pm 6.43 \\
			93.52 \pm 4.47\end{array}$ & $\begin{array}{l}89.61 \pm 7.89 \\
			90.77 \pm 6.70\end{array}$ & 200 & 16,000 \\
		\cline{2-8} & FedProx \cite{li2020federated} & $\begin{array}{l}1 \\
			2\end{array}$ & $\begin{array}{l}94.53 \pm 5.33 \\
			93.49 \pm 5.30\end{array}$ & $\begin{array}{l}90.71 \pm 6.24 \\
			93.74 \pm 5.02\end{array}$ & $\begin{array}{l}91.33 \pm 7.32 \\
			89.49 \pm 6.74\end{array}$ & 300 & 16,000 \\
		\cline{2-8} & Fedper \cite{arivazhagan2019federated} & $\begin{array}{l}1 \\
			2\end{array}$ & $\begin{array}{l}93.47 \pm 5.44 \\
			92.27 \pm 6.16\end{array}$ & $\begin{array}{l}90.22 \pm 7.63 \\
			91.99 \pm 6.33\end{array}$ & $\begin{array}{l}87.73 \pm 9.64 \\
			87.54 \pm 8.14\end{array}$ & 250 & 102 \\
		\cline{2-8} & FedAvg \cite{mcmahan2017communication} & $\begin{array}{l}1 \\
			2\end{array}$ & $\begin{array}{l}94.50 \pm 5.29 \\
			94.13 \pm 4.92\end{array}$ & $\begin{array}{l}91.39 \pm 5.23 \\
			93.02 \pm 5.77\end{array}$ & $\begin{array}{l}90.95 \pm 7.22 \\
			89.80 \pm 6.94\end{array}$ & 300 & 16,000 \\
		\cline{2-8} & FedRep \cite{collins2021exploiting}& $\begin{array}{l}1 \\
			2\end{array}$ & $\begin{array}{l}93.36 \pm 5.34 \\
			92.28 \pm 5.40\end{array}$ & $\begin{array}{l}91.41 \pm 5.89 \\
			91.56 \pm 7.02\end{array}$ & $\begin{array}{l}89.98 \pm 6.88 \\
			88.23 \pm 6.97\end{array}$ & 200 & 102 \\
		\cline{2-8} & FedProto \cite{tan2022fedproto} & $\begin{array}{l}1 \\
			2\end{array}$ & $\begin{array}{l}96.82 \pm 1.75 \\
			94.93 \pm 1.29\end{array}$ & $\begin{array}{l}94.93 \pm 1.61 \\
			\textbf{94.69} \pm \textbf{1.50}\end{array}$ & $\begin{array}{l}93.67 \pm 2.23 \\
			93.03 \pm 2.50\end{array}$ & 120 & 4 \\
		\cline{2-8} & \textbf{FedRFQ} & $\begin{array}{l}1 \\
			2\end{array}$ & $\begin{array}{l}\textbf{97.76} \pm \textbf{1.68} \\
			\textbf{96.17} \pm \textbf{2.23}\end{array}$ & $\begin{array}{l}\textbf{95.08} \pm \textbf{2.15} \\
			94.56 \pm 1.93\end{array}$ & $\begin{array}{l}\textbf{93.77} \pm \textbf{2.03} \\
			\textbf{94.53} \pm \textbf{1.57} \end{array}$ & 120 & \textbf{1.92} \\
		\hline
	\end{tabular}
\label{table2}
\end{table*}
\begin{table*}[!htb]
\centering
\caption{Comparison of federated learning methods on the CIFAR-10 dataset under non-IID settings. The experimental results for FedRFQ come from the following parameter settings  $\eta=0.01, \lambda=0.1, \zeta=0, \psi=0$ since these values can achieve better results through our continuous test. The best results are shown in bold. The results indicate that FedRFQ achieves higher accuracy with fewer communication parameters.}
\renewcommand\arraystretch{1.03}
\begin{tabular}{|c|c|c|c|c|c|c|c|}
	\hline \multirow{2}{*}{ Dataset } & \multirow{2}{*}{ Method } & \multirow{2}{*}{$\begin{array}{c}\mathsf{std} \\
			\text { of } \mathsf{avg}\end{array}$} & \multicolumn{3}{|c|}{ Test Accuracy } & \multirow{2}{*}{$\begin{array}{c}\text {Communication } \\
			\text { Rounds }\end{array}$} & \multirow{2}{*}{$\begin{array}{c}\text {Communication } \\
			\text { Params }\left(\times 10^4\right)\end{array}$} \\
	\cline{4-6} & & & $\mathsf{avg}=3$ & $\mathsf{avg}=4$ & $\mathsf{avg}=5$ & & \\
	\hline \multirow{14}{*}{CIFAR-10} & Local & $\begin{array}{l}1 \\
		2\end{array}$ & $\begin{array}{l}79.72 \pm 9.45 \\
		68.15 \pm 9.88\end{array}$ & $\begin{array}{c}67.62 \pm 7.15 \\
		61.03 \pm 11.83\end{array}$ & $\begin{array}{c}58.64 \pm 6.57 \\
		58.81 \pm 12.90\end{array}$ & 0 & 0 \\
	\cline{2-8} & FeSEM \cite{xie2021multi} & $\begin{array}{l}1 \\
		2\end{array}$ & $\begin{array}{l}80.19 \pm 3.31 \\
		76.12 \pm 4.15\end{array}$ & $\begin{array}{l}76.40 \pm 3.23 \\
		72.11 \pm 3.48\end{array}$ & $\begin{array}{l}74.17 \pm 3.51 \\
		70.89 \pm 3.39\end{array}$ & 120 & $2.35 \times 10^4$ \\
	\cline{2-8} & FedProx \cite{li2020federated} & $\begin{array}{l}1 \\
		2\end{array}$ & $\begin{array}{l}83.25 \pm 2.44 \\
		79.83 \pm 2.35\end{array}$ & $\begin{array}{l}79.20 \pm 1.31 \\
		72.56 \pm 1.90\end{array}$ & $\begin{array}{l}76.19 \pm 2.23 \\
		71.39 \pm 2.36\end{array}$ & 150 & $2.35 \times 10^4$ \\
	\cline{2-8} & FedPer \cite{arivazhagan2019federated} & $\begin{array}{l}1 \\
		2\end{array}$ & $\begin{array}{l}84.38 \pm 4.58 \\
		\textbf{84.51} \pm \textbf{4.39}\end{array}$ & $\begin{array}{l}78.73 \pm 4.59 \\
		73.31 \pm 4.76\end{array}$ & $\begin{array}{l}76.21 \pm 4.27 \\
		72,43 \pm 4.55\end{array}$ & 130 & $2.25 \times 10^4$ \\
	\cline{2-8} & FedAvg \cite{mcmahan2017communication} & $\begin{array}{l}1 \\
		2\end{array}$ & $\begin{array}{l}81.72 \pm 2.77 \\
		78.99 \pm 2.34\end{array}$ & $\begin{array}{l}76.77 \pm 2.37 \\
		72.73 \pm 2.58\end{array}$ & $\begin{array}{l}75.74 \pm 2.61 \\
		70.93 \pm 2.82\end{array}$ & 150 & $2.35 \times 10^4$ \\
	\cline{2-8} & FedRep \cite{collins2021exploiting} & $\begin{array}{l}1 \\
		2\end{array}$ & $\begin{array}{l}81.44 \pm 10.48 \\
		76.70 \pm 11.79\end{array}$ & $\begin{array}{c}76.93 \pm 7.46 \\
		73.54 \pm 11.42\end{array}$ & $\begin{array}{l}73.36 \pm 7.04 \\
		70.30 \pm 8.00\end{array}$ & 110 & $2.25 \times 10^4$ \\
	\cline{2-8} & FedProto \cite{tan2022fedproto} & $\begin{array}{l}1 \\
		2\end{array}$ & $\begin{array}{l}8 4 . 4 9 \pm 1 . 9 7 \\
		81.75 \pm 1.39\end{array}$ & $\begin{array}{l}79.12 \pm 2.03 \\
		74.98 \pm 1.61\end{array}$ & $\begin{array}{l}77.08 \pm 1.98 \\
		71.17 \pm 1.29\end{array}$ & 110 & 4.10 \\
	\cline{2-8} & \textbf{FedRFQ} & $\begin{array}{l}1 \\
		2\end{array}$ & $\begin{array}{l}\textbf{85.99} \pm \textbf{1.59} \\
		83.82 \pm 1.93\end{array}$ & $\begin{array}{l}\textbf{82.02} \pm \textbf{1.98} \\
		\textbf{81.53} \pm \textbf{1.28}\end{array}$ & $\begin{array}{l}\textbf{79.84} \pm \textbf{1.48} \\
		\textbf{79.47} \pm \textbf{1.40}\end{array}$ & 110 & \textbf{1.97} \\
	\hline
\end{tabular}
\label{table3}
\end{table*}
\begin{figure*}[!htb]
	\hspace{-0.6cm}
	\begin{minipage}[t]{0.55\linewidth}
		\centering
		\includegraphics[width=3.5in,height=1.9in]{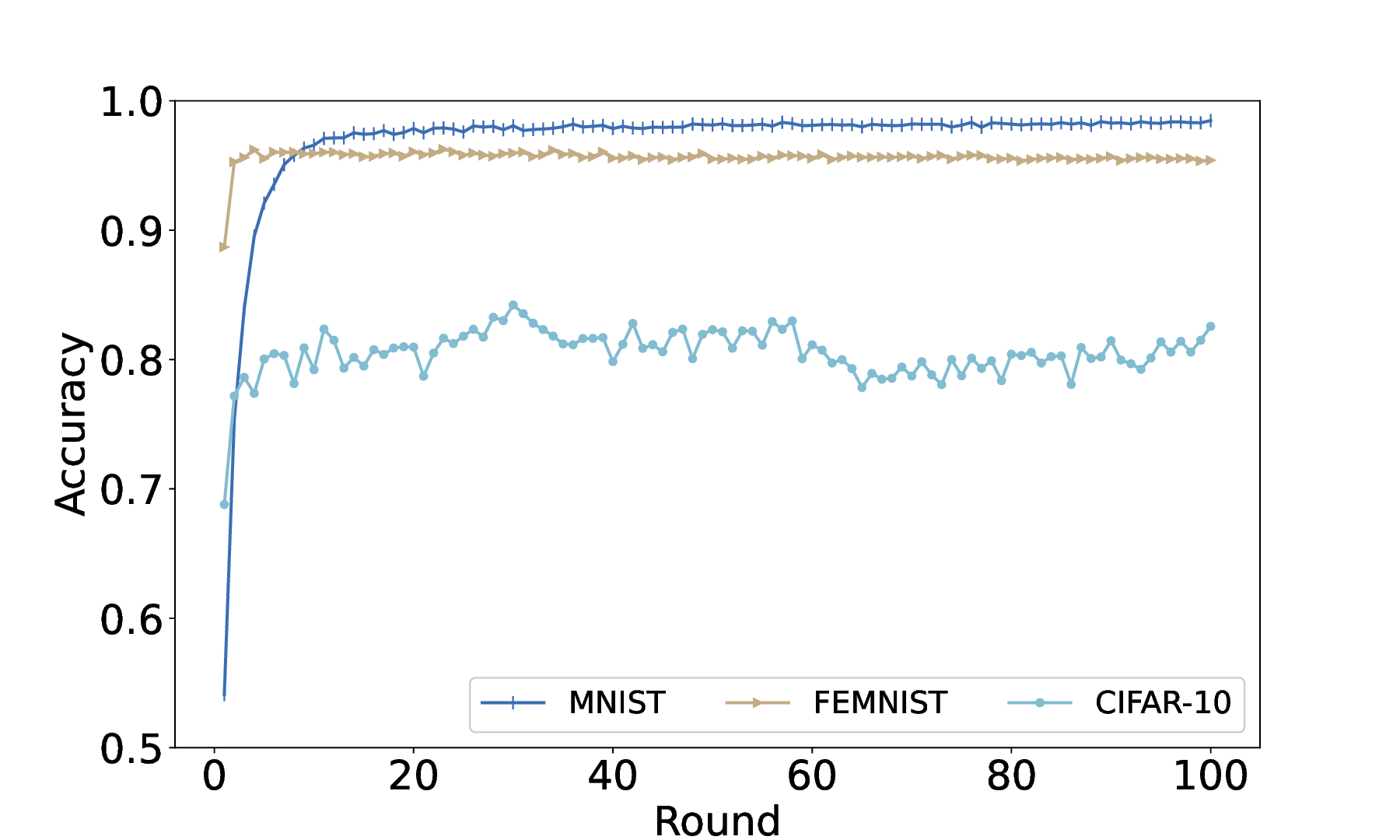}
	\end{minipage}
	\hspace{-0.9cm}
	\begin{minipage}[t]{0.55\linewidth}
		\centering
		\includegraphics[width=3.5in,height=1.9in]{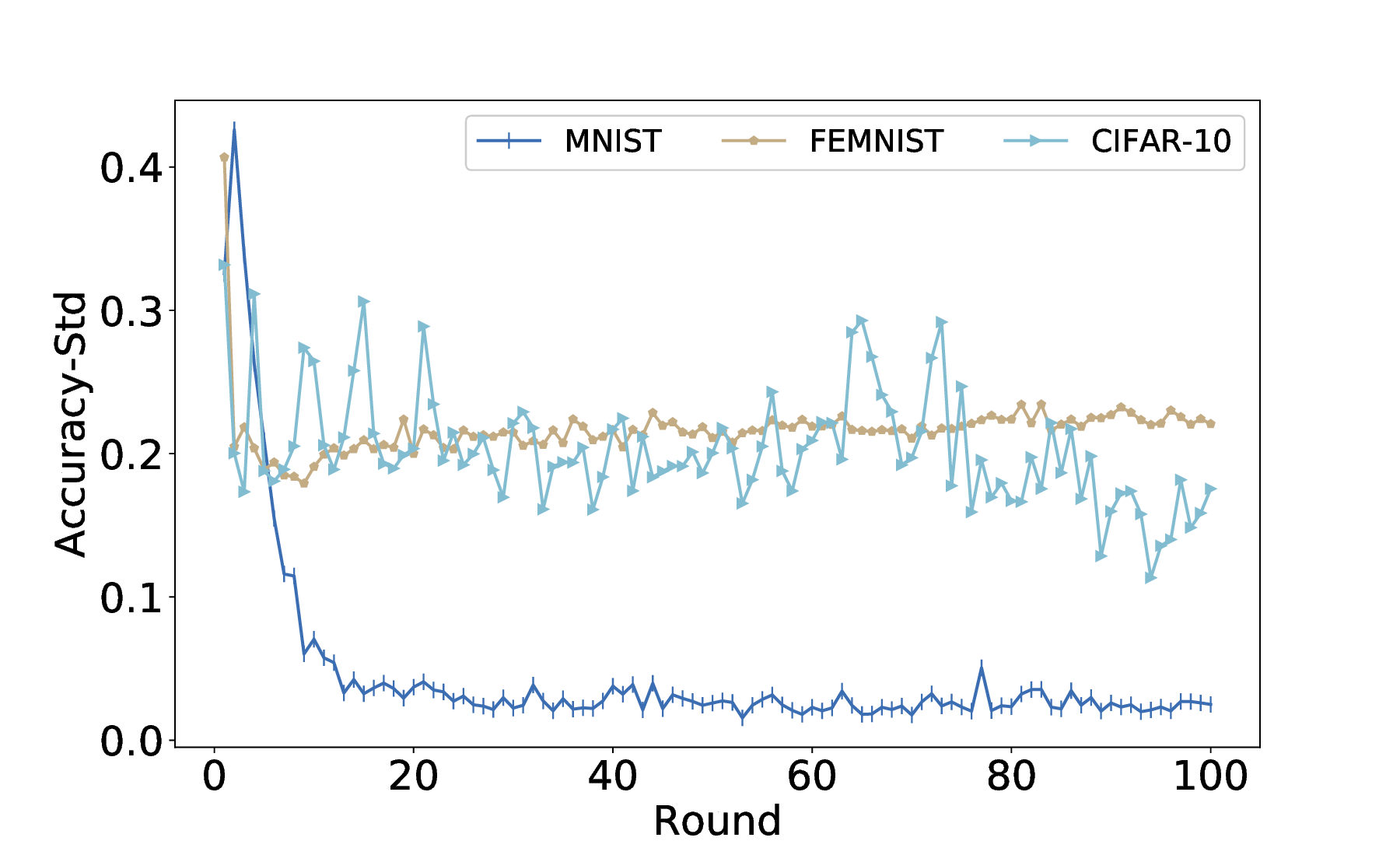}
	\end{minipage}
	\caption{The accuracy and its standard deviation of FedRFQ versus the total number of communication rounds for MNIST, FEMNIST, and CIFAR-10 under non-IID, where $\mathsf{avg} = 3, \mathsf{std} = 2, \eta=0.01, \lambda=1, \zeta=0, \psi=0$.}
	\label{qdmq}
\end{figure*}
\begin{figure}[!htb]
	\centering \includegraphics[width=3.5in,height=1.9in]{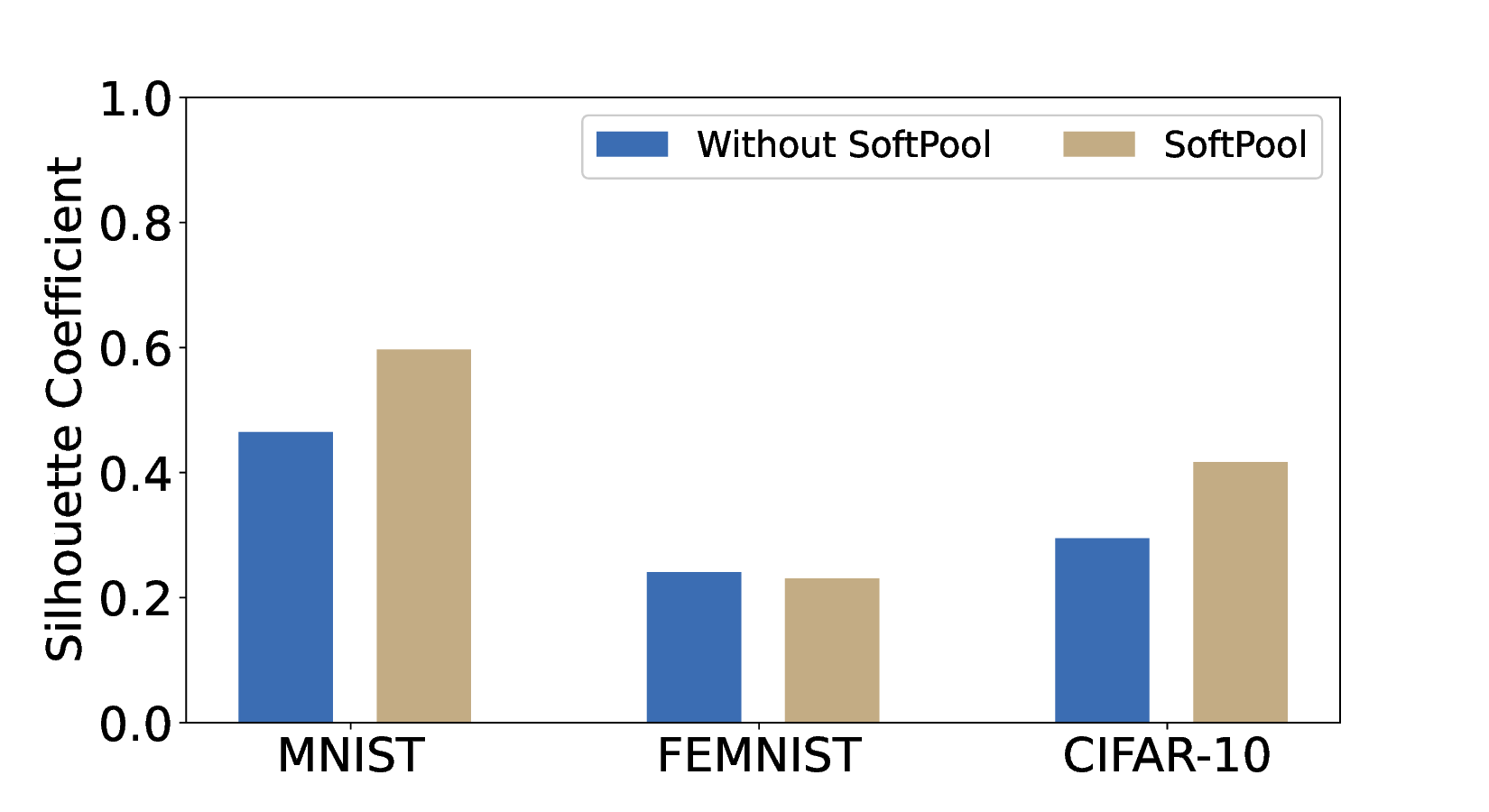}
	\caption{The clustering effects of prototypes in FedRFQ with and without SoftPool applied on the MNIST, FEMNIST, and CIFAR-10 datasets.}
	\label{tre1}
\end{figure}

\section{Evaluation}\label{experiments}
In this section, we introduce the experimental settings, report 
 our performance evaluation results, and analyze the security properties of FedRFQ under various learning parameters.

\subsection{Settings}
\textit{1) Datasets and Models}: In our experiments, we use three datasets (MNIST, FEMNIST, CIFAR-10) considering non-IID settings to evaluate FedRFQ. The local model for the MNIST and FEMNIST datasets employs a multi-layer Convolutional Neural Network (CNN) architecture that comprises $2$ convolutional layers and $2$ fully-connected layers; and the local model for the CIFAR-10 dataset is the ResNet18 model architecture. It is worth noting that the ResNet18 model employed by all clients is pre-trained on the ImageNet dataset and serves as the initial local model. 

\textit{2) Baselines}: To evaluate FedRFQ's performance over non-IID data, we compare it with the following baselines: Local,  FeSEM \cite{xie2021multi}, FedProx \cite{li2020federated}, FedPer \cite{arivazhagan2019federated}, FedAvg \cite{mcmahan2017communication}, FedRep \cite{collins2021exploiting} and FedProto \cite{tan2022fedproto}. Note that the Local model is trained for each client individually without communicating with other clients.

\textit{3) Experiment Details}: We use PyTorch to conduct the experiments for FedRFQ and all baseline methods. In the non-IID setting, we assume that the statistical distributions of the clients are diverse. To simulate differences in the degree of non-IID, we denote the average number of classes and its standard deviation in all clients as $\mathsf{avg}$ and $\mathsf{std}$, respectively. In the collaborative training, we use a consortium of $20$ clients, with each being trained with a SGD optimizer. Each client is randomly configured with a certain number of classes, ensuring that $\mathsf{avg}$ is either $3$, $4$, or $5$, and $\mathsf{std}$ is either $1$, $2$, or $3$. In any client, each class contains the same number of samples, which are assigned randomly. By doing so, we can create heterogeneity of samples that simulate the non-IID data present on clients in real-world scenarios.

\subsection{The Performance of FedRFQ}
Our results indicate that FedRFQ improves the accuracy of the local models under the non-IID settings. Table~\ref{table1}, Table~\ref{table2}, and Table~\ref{table3} report the accuracy of different methods with various parameters for MNIST, FEMNIST, and CIFAR-10 under non-IID \cite{tan2022fedproto}. It can be seen that FedRFQ achieves the highest accuracy in most cases, which indicate that FedRFP can effectively mitigate the challenges of non-IID. Compared with other methods, FedRFQ converges faster for local optimization in most cases, as it takes fewer number of communication rounds for the three datasets. Moreover, it also demonstrates that more communication parameters do not necessarily lead to better results under non-IID. Hence, the accuracy of FedRFQ relies not only on the number of communication parameters, but also on the crucial information of these parameters.

\begin{figure}[!htb]
\centering 
\includegraphics[width=3.5in,height=1.9in]{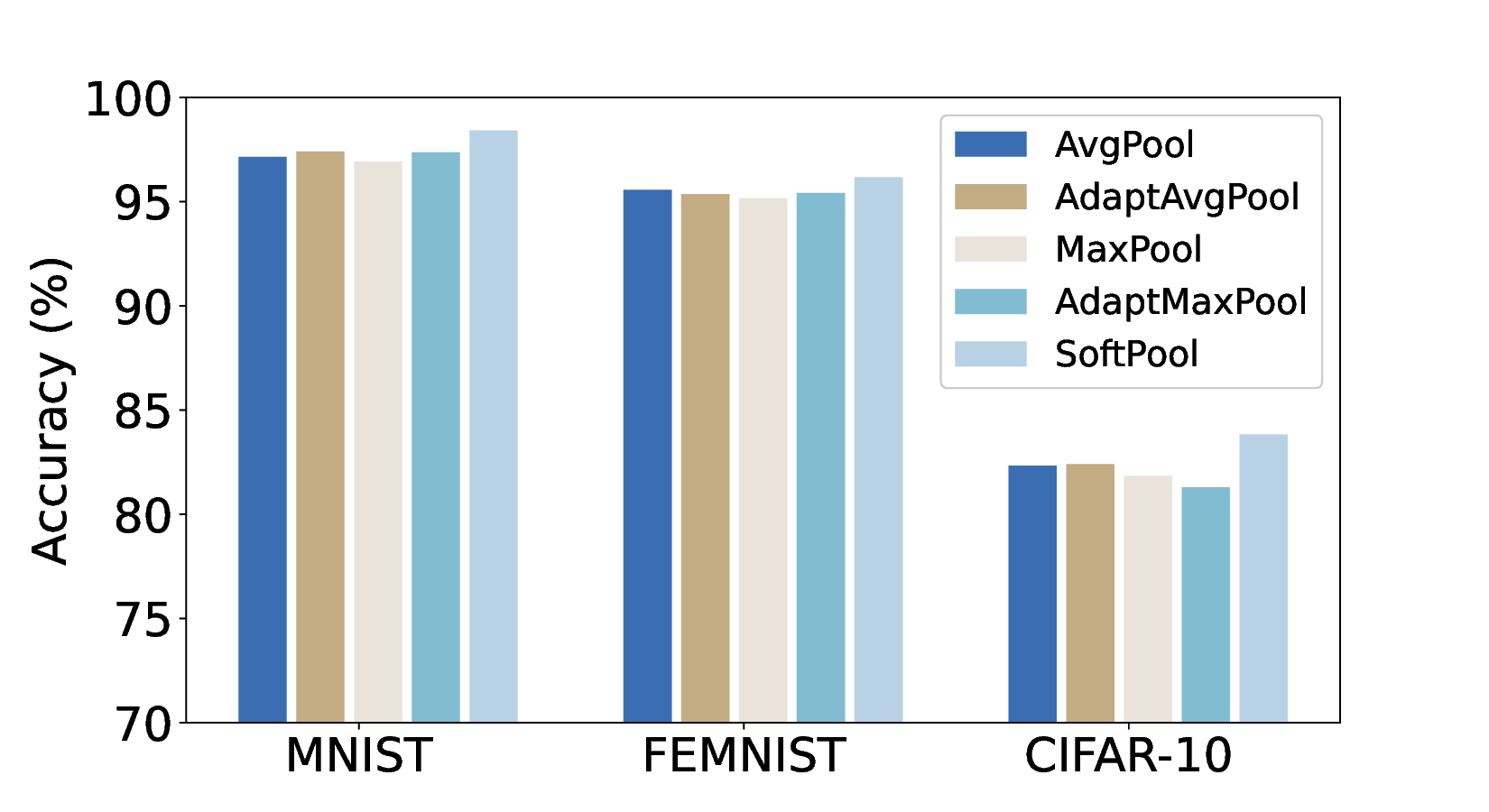}
\caption{The accuracy of the local models for all clients with different pooling methods for MNIST, FEMNIST, and CIFAR-10 under non-IID, where $\mathsf{avg} = 3$, $\mathsf{std} = 2$, $\eta=0.01$, $\lambda=1$, $\zeta=0$, $\psi=0$, $\mathsf{round} = 100$. }
\label{atao}
\end{figure}

Fig.~\ref{qdmq} shows the accuracy and its standard deviation on MNIST, FEMNIST, and CIFAR-10 under non-IID over different communication rounds. One can see that for MNIST and FEMNIST, the accuracy is high and stable. In contrast, CIFAR-10 demonstrates lower accuracy with significant fluctuations. This discrepancy can be attributed to the inherent complexity of the datasets. MNIST and FEMNIST are simpler, which allow each client to achieve high accuracy relatively easily with minimal variance across different clients. On the contrary, the complexity of CIFAR-10 leads to more pronounced variations in accuracy among clients. Consequently, FedRFQ exhibits optimal performance for MNIST but is less effective for CIFAR-10.

FedRFQ reduces the differences between prototypes corresponding to samples within the same class and makes these prototypes closer to the objective prototypes of the respective classes, further improving the accuracy of the local model. We use unsupervised clustering to measure the reduction in prototype differences for samples belonging to the same class. The evaluation of the clustering effect is conducted using the silhouette coefficient. Typically, a higher silhouette coefficient value indicates better clustering among prototypes and clearer boundaries between different clusters. 
In experiments conducted on the MNIST, FEMNIST, and CIFAR-10 datasets under non-IID settings, Fig.~\ref{tre1} illustrates the silhouette coefficient of the prototypes on test data under two different conditions: with and without the utilization of the SoftPool operation. The results demonstrate that the use of SoftPool in the MNIST and CIFAR-10 datasets significantly enhances the silhouette coefficient of prototypes, indicating a better clustering effect. However, on the FEMNIST dataset, the influence of SoftPool on the silhouette coefficient is constrained. This limitation arises from the large number of classes in FEMNIST, which leads to a close distribution of prototypes across different classes. Consequently, this close distribution diminishes the clustering effect with SoftPool. In summary, FedRFQ with SoftPool reduces the dissimilarity between prototypes of samples within the same class compared to that without SoftPool.

\begin{figure*}[!htb]  
  \centering
    \subfigure[MNIST]{
    \includegraphics[width=0.31\linewidth]{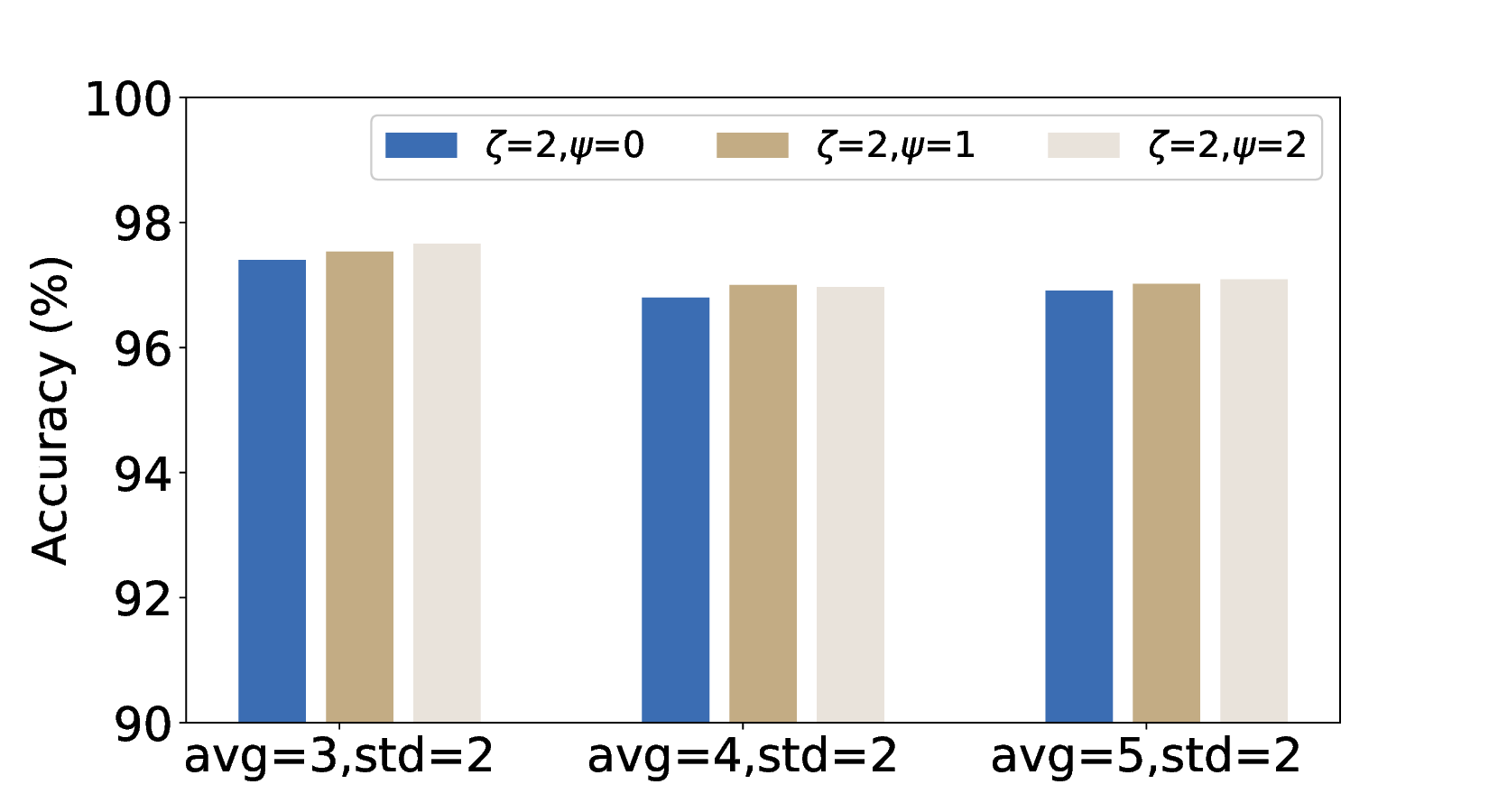}
    }
    \subfigure[FEMNIST]{
    \includegraphics[width=0.31\linewidth]{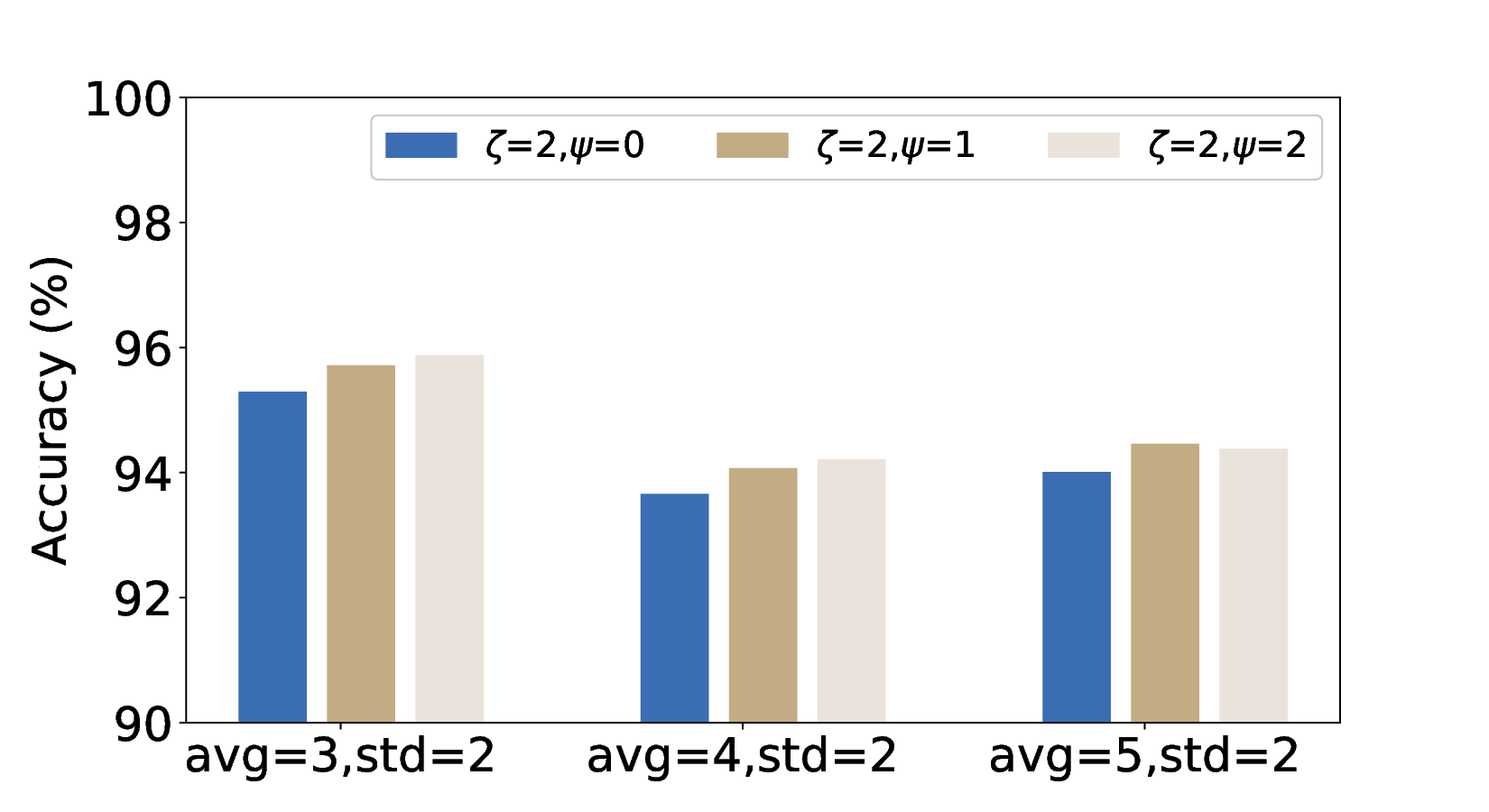}
    }
    \subfigure[CIFAR-10]{
    \includegraphics[width=0.31\linewidth]{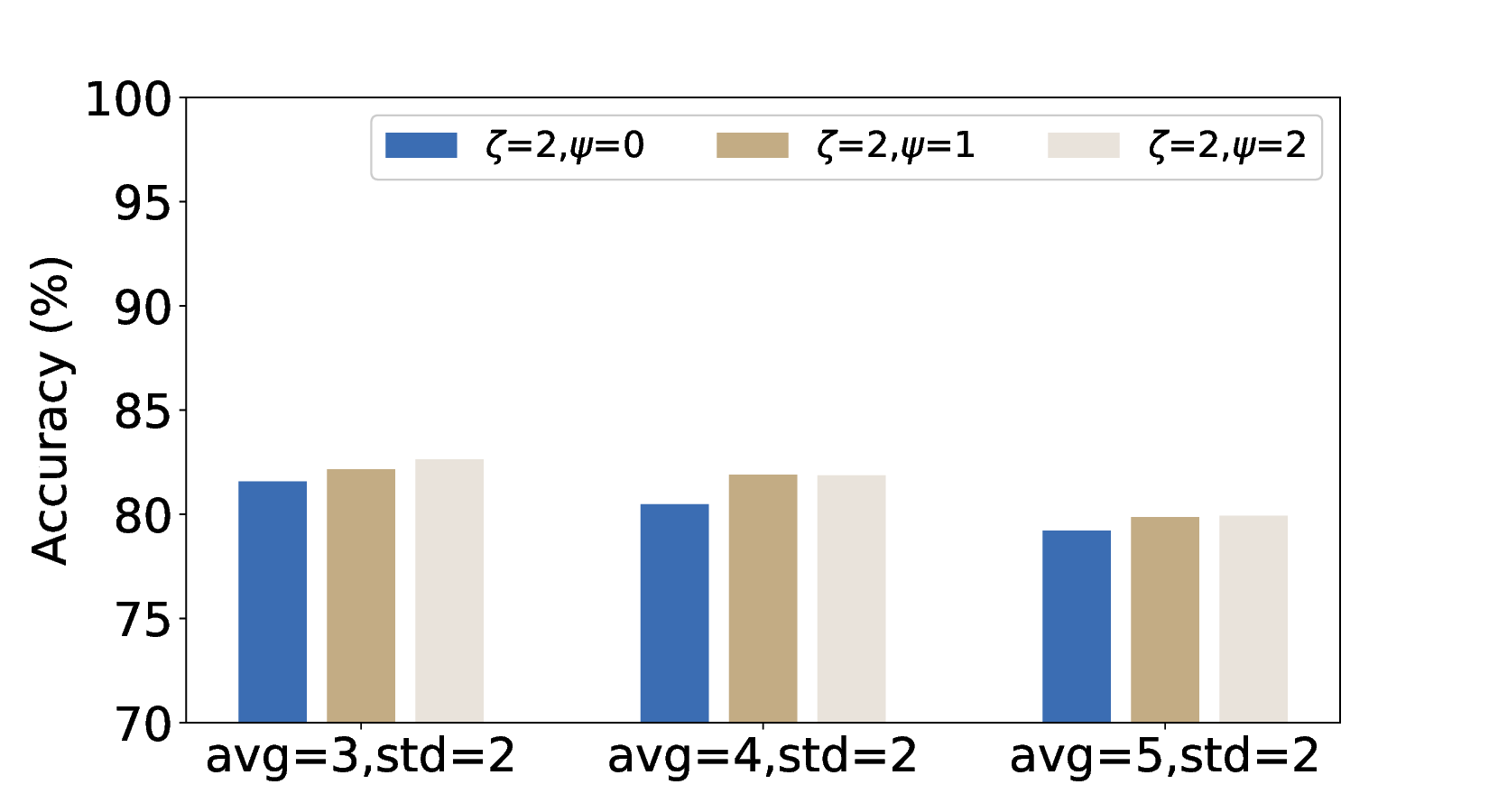}
    }
  \caption{The accuracy of FedRFQ versus degrees of non-IID with poisoning attacks on MNIST, FEMNIST, and CIFAR-10, where $ \mathsf{round} = 100, \eta=0.01, \lambda=1$.}
  \label{qdms}
\end{figure*}
\begin{figure}[!htb]
	\includegraphics[width=3.5in,height=1.9in]{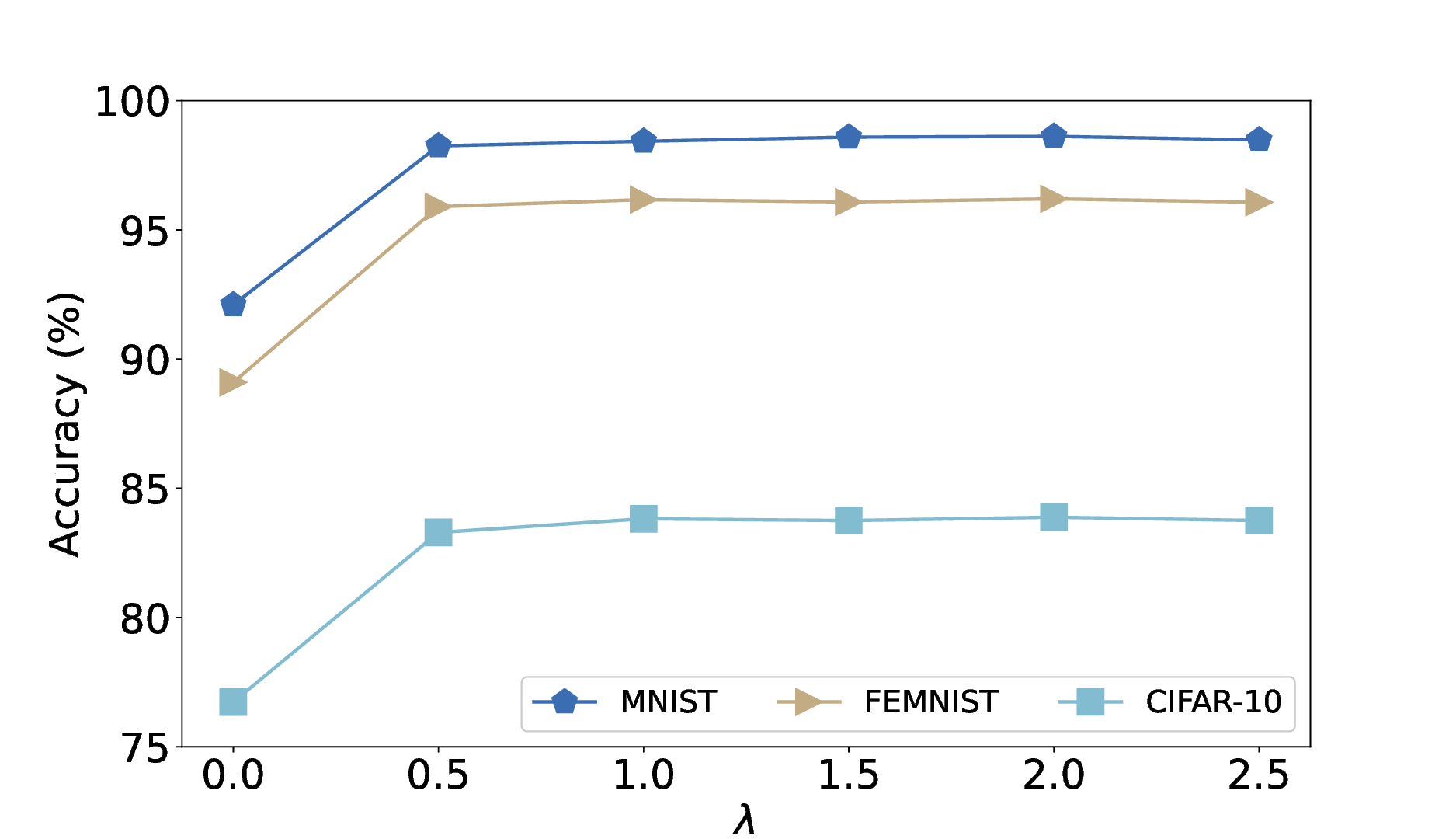}
	\caption{The accuracy of  FedRFQ versus $\lambda$  on  MNIST, FEMNIST, and CIFAR-10 under non-IID, where $\mathsf{avg} = 3, \mathsf{std} = 2, \mathsf{round} = 100, \eta=0.01$, $\zeta=0, \psi=0$.}
	\label{tre}
\end{figure}

In order to validate the superiority of SoftPool for prototype enhancement in FedRFQ, we introduce other pooling methods for comparison, such as AvgPool, MaxPool, AdaptAvgPool, AdaptMaxPool and SoftPool. The experimental results, depicted in Fig.~\ref{atao}, demonstrate the performance of different pooling methods on MNIST, FEMNIST and CIFAR-10. One can see that each pooling method exhibits varying accuracy across datasets. However, SoftPool consistently outperforms other pooling methods, particularly in mitigating differences between prototypes of samples from the same class.
AvgPool and MaxPool ignore the significance of local minimum values within the kernel region, which damages the important features of the prototype. Although AdaptAvgPool and AdaptMaxPool integrate a self-adaptation mechanism, they still rely on average or maximum values within the kernel region, leading to a loss of prototype features. In summary, SoftPool exhibits superior performance compared to other pooling methods. It fully considers scenarios with heterogeneous data in federated learning and generates a high-quality global prototype. Furthermore, the above experimental results demonstrate that embedding SoftPool into FedRFQ has a significant positive impact summarized as follows.
\begin{itemize}
	\item[1)] Reducing prototype redundancy: SoftPool can reduce the number of prototypes that need to be transmitted because it can compress the prototype vectors by an exponentially weighted average.
	\item[2)] Improving aggregation performance on non-IID: The SoftPool can eliminate the distance between prototypes of the same class across different clients, which improves the capability of handling non-IID data.
\end{itemize}

\subsection{Poisoning Attacks on FedRFQ} 
In order to simulate the impact of malicious clients, a subset of clients is designated as malicious, denoted by $\zeta$. Fig.~\ref{qdms} illustrates the experimental results of FedRFQ with poisoning attacks on MNIST, FEMNIST and CIFAR-10 under the non-IID settings. The number of malicious clients is uniformly set as $\zeta = 2$, and the security level is denoted as $\psi$. Notably, the accuracy of FedRFQ is observed to be higher when the security level $\psi=1$ or $2$ compared to that of $\psi=0$. This experimental result indicates that BFT-detect can effectively resist poisoning attacks orchestrated by malicious clients. We observe that their accuracy is comparable when $\psi=1$ or $2$, and both outperform the accuracy when $\psi=0$, emphasizing the effectiveness of FedRFQ in resisting poisoning attacks.

\subsection{The Impact of Parameter $\lambda$ on FedRFQ} 
The experimental results depicted in Fig.~\ref{tre} show the performance of FedRFQ on FEMNIST, MNIST and CIFAR-10 under the non-IID settings with varying values of the parameter $\lambda$. When $\lambda = 0$, the accuracy of FedRFQ is the lowest as FedRFQ is trained locally without communication with other clients. As each client is trained locally and independently without prototype information exchanges, it cannot benefit from federated training. However, as the value of $\lambda$ gradually increases, the importance of global prototype information increases and the accuracy of FedRFQ is significantly improved. Note that once the value of $\lambda$ surpasses $0.5$, the increase of $\lambda$ does not obviously improve the accuracy, which suggests that the significance of prototype information has reached a stable level. In summary, this result indicates that the interaction of prototype information can improve the effectiveness of the local model.

\section{conclusion}\label{conclusion}
FedRFQ introduces an innovative prototype-based federated learning framework designed to effectively alleviate challenges associated with prototype redundancy and prototype failure, concurrently providing resilience against poisoning attacks and server malfunctions. Specifically, FedRFQ employs SoftPool to tackle issues related to prototype redundancy and prototype failure. Additionally, a novel BFT detectable aggregation algorithm, denoted as BFT-detect, is introduced to ensure the security of FedRFQ. BFT-detect can resist poisoning attacks from malicious clients and avoid server malfunctions. This study includes a comprehensive analysis that evaluates the effectiveness of FedRFQ. Extensive experiments are conducted, demonstrating the superior performance of FedRFQ compared to other state-of-the-art methods.

\section{Acknowledgement}
This study was partially supported by the National Key R\&D Program of China (No.2022YFB4501000), the National Natural Science Foundation of China (No.62232010, 62302266, U23A20302), Shandong Science Fund for Excellent Young Scholars (No.2023HWYQ-008), Shandong Science Fund for Key Fundamental Research Project (ZR2022ZD02), and the Fundamental Research Funds for the Central Universities.

\ifCLASSOPTIONcaptionsoff
  \newpage
\fi



%
\bibliographystyle{IEEEtran}
\bibliography{FedRFQ.bib}

%



\onecolumn
\begin{appendices}
\section{The proof of Theorm 1} \label{theorem111}
\begin{lemma}
Assume that \textit{Assumption} \ref{assumption1} and \ref{assumption2} hold, we can establish the following bound on the local loss function of any given client from the start of the $(t + 1)$-th round communication to the completion of the final step:
\begin{footnotesize}
\begin{equation}
	\begin{aligned}
		\mathbb{E}\left[\bm{\ell}_k^{(t+1)E}\right] \leq &\bm{\ell}_k^{(tE)^\prime}-\left(\eta-\frac{L_a \eta^2}{2}\right)( \sum_{e=1}^{E-1}\left\|\nabla \bm{\ell}_k^{tE+e}\right\|_2^2 + \left\|\nabla \bm{\ell}_k^{(t E)^\prime}\right\|_2^2)+\frac{L_a E \eta^2}{2} \sigma^2. 
	\end{aligned}
\nonumber
\end{equation}
\end{footnotesize}
\label{lemma1}
\end{lemma}

\begin{IEEEproof}
In Theorem \ref{theorem1}, the local loss function of each client is consistent, therefore, to simplify the proof, the index $k$ can be omitted, which does not affect the proof. Consider the iterations of federated learning at $(tE)^\prime$ and $tE+1$, the local loss functions at these two iterations are $\bm{\ell}^{(tE)^\prime}$ and $\bm{\ell}^{tE+1}$, respectively. According to the  $L_a$-Lipschitz smooth bound in \textit{Assumption} \ref{assumption1}, we have
\begin{footnotesize}
\begin{equation}
\begin{aligned}
\bm{\ell}^{tE+1} \leq & \bm{\ell}^{(tE)^\prime}+\left\langle\nabla \bm{\ell}^{(tE)^\prime},\left(\omega^{tE+1}-\omega^{(tE)^\prime}\right)\right\rangle+ \frac{L_a}{2}\left\|\omega^{tE+1}-\omega^{(tE)^\prime}\right\|_2^2.
\end{aligned}
\label{ee2}
  \end{equation}
\end{footnotesize}

According to the gradient update principle: $\omega^{t+1}=\omega^t-\eta g^t$, we have $\omega^{tE+1}=\omega^{(tE)^\prime}-\eta g^{(tE)^\prime}$, which is substituted into Eq. (\ref{ee2}) and we can obtain Eq. (\ref{17}). 
\begin{footnotesize}
\begin{equation}
\bm{\ell}^{tE+1} \leq \bm{\ell}^{(tE)^\prime}-\eta\left\langle\nabla \bm{\ell}^{(tE)^\prime}, g^{(tE)^\prime}\right\rangle+\frac{L_a}{2}\left\|\eta g^{(tE)^\prime}\right\|_2^2. 
\label{17}
\end{equation}
\end{footnotesize}
\begin{footnotesize}
\begin{equation}
\begin{aligned}
\mathbb{E}\left[\bm{\ell}^{tE+1}\right] & \leq \bm{\ell}^{(tE)^\prime}-\eta \mathbb{E}\left[\left\langle\nabla \bm{\ell}^{(tE)^\prime}, g^{(tE)^\prime}\right\rangle\right]+\frac{L_a \eta^2}{2} \mathbb{E}\left[\left\|g^{(tE)^\prime}\right\|_2^2\right] \\
&\stackrel{(s1)}{=} \bm{\ell}^{(tE)^\prime}-\eta \mathbb{E}\left[\left\langle\nabla \bm{\ell}^{(tE)^\prime}, \nabla \bm{\ell}^{(tE)^\prime}\right\rangle\right]+\frac{L_a \eta^2}{2} \mathbb{E}\left[\left\|g^{(tE)^\prime}\right\|_2^2\right] \\
& = \bm{\ell}^{(tE)^\prime}-\eta\left\|\nabla \bm{\ell}^{(tE)^\prime}\right\|_2^2+\frac{L_a \eta^2}{2} \mathbb{E}\left[\left\|g^{(tE)^\prime}\right\|_2^2\right] \\
& \stackrel{(s2)}{=} \bm{\ell}^{(tE)^\prime}-\eta\left\|\nabla \bm{\ell}^{(tE)^\prime}\right\|_2^2+\frac{L_a \eta^2}{2}\left((\mathbb{E}\left[g^{(tE)^\prime}\right])^2+\operatorname{Var}\left(g^{(tE)^\prime}\right)\right) \\
&\stackrel{(s3)} {=} \bm{\ell}^{(tE)^\prime}-\eta\left\|\nabla \bm{\ell}^{(tE)^\prime}\right\|_2^2+\frac{L_a \eta^2}{2}\left(\left\|\nabla \bm{\ell}^{(tE)^\prime}\right\|_2^2+\operatorname{Var}\left(g^{(tE)^\prime}\right)\right) \\
& =\bm{\ell}^{(tE)^\prime}-\left(\eta-\frac{L \eta^2}{2}\right)\left\|\nabla \bm{\ell}^{(tE)^\prime2}\right\|_2^2+\frac{L_a \eta^2}{2} \operatorname{Var}\left(g^{(tE)^\prime}\right) \\
& \stackrel{(s4)} {\leq} \bm{\ell}^{(tE)^\prime}-\left(\eta-\frac{L \eta^2}{2}\right)\left\|\nabla \bm{\ell}^{(tE)^\prime}\right\|_2^2+\frac{L_a \eta^2}{2} \sigma^2. \label{18}
\end{aligned}
\end{equation}
\end{footnotesize}

The $\bm{\ell}^{(tE)^\prime}$  is a fixed value to calculate the bound of $\bm{\ell}^{tE+1}$, that is, we calculate the expectation on both sides of (\ref{17}) with different samples. After a series of transformations, we can get Eq. (\ref{18}), in which $(s1)$ is based on \textit{Assumption} \ref{assumption2}, $(s2)$ is based on $\mathtt{Var}(x)=\mathbb{E}\left[x^2\right]-(\mathbb{E}[x])^2$; we can also get that $\mathtt{Var}(g^{(tE)^\prime})=\mathbb{E}\left[||g^{(tE)^\prime}||_2^2\right]-(\mathbb{E}[g^{(tE)^\prime}])^2$, where $\mathtt{Var}(g^{(tE)^\prime})$ denotes the variance of random variable $g^{(tE)^\prime}$. Note that $(s3)$ is also based on \textit{Assumption} \ref{assumption2}, and $(s4)$ is based on Eq. (\ref{ee3}) in \textit{Assumption} \ref{assumption2} and it is also based on the assumption that the variance of $g^{(tE)^\prime}$ is bounded by $\sigma^2$.
After $E$ iterations of Eq. (\ref{18}), the bound of the local loss function is Eq. (\ref{shang}),
\begin{footnotesize}
\begin{equation}
\begin{aligned}
\mathbb{E}\left[\bm{\ell}^{(t+1)E}\right] \leq & \bm{\ell}^{(tE)^\prime}-\left(\eta-\frac{L_a \eta^2}{2}\right) (\sum_{e=1}^{E-1}\left\|\nabla \bm{\ell}^{tE+e}\right\|_2^2+\left\|\nabla \bm{\ell}^{(tE)^\prime}\right\|_2^2) +\frac{L_a E \eta^2}{2} \sigma^2,
\label{shang}
\end{aligned}
\end{equation}
\end{footnotesize}
which means that the boundary range of any client's loss function can be expressed by the Eq. (\ref{shang}) from the beginning to the end of the ($t+1$)-th round communication.
\end{IEEEproof}

\begin{lemma}
Assume that \textit{Assumption} \ref{assumption3} and \ref{assumption4} hold. After prototype aggregation at the servers, the loss function of an arbitrary client can be bounded as:
\begin{footnotesize}
\begin{equation}
	\begin{aligned}
		\mathbb{E}[\bm{\ell}^{((t+1)E)^\prime}] &\leq \bm{\ell}^{(t+1)E}+\lambda L_{2} L_{2}^{\prime} \eta \sum_{k=1}^K v_k (\sum_{e=1}^{E-1} \mathbb{E}[\| g_{k}^{tE+e} \|_2]+\mathbb{E}[\| g_{k}^{(tE)^\prime} \|_2])+ 2\varepsilon \\
		&\leq \bm{\ell}^{(t+1)E}+\lambda L_{2} L_{2}^{\prime} \eta E G + 2\varepsilon. \\
		\nonumber
	\end{aligned}
\end{equation}
\end{footnotesize}
\label{lemma2}
\end{lemma}
\begin{IEEEproof}
After a client is trained for $(t+1)E$ iteration, the loss function state $\bm{\ell}^{(t+1)E}$ of the client is obtained. Then, one aggregates the local prototypes in the servers and obtains the loss function state $\bm{\ell}^{((t+1)E)^\prime}$ of the client.

Assume that $\bm{\ell}^{((t+1)E)^\prime}=\bm{\ell}^{(t+1)E}+\bm{\ell}^{((t+1)E)^\prime}-\bm{\ell}^{(t+1)E}$. The reasoning process of the equation is shown in Eq. (\ref{21}). Further, we  combine Eq. (\ref{lof88}) and simplify the loss of the local prototype $\sum^J_{j} \left\|P_k^{(j)}-\mathcal{P}^{(j)}\right\|_2$ as $\left\|P-\mathcal{P} \right\|_{2}$, to get $(s5)$. Then, since the weights of the local models at $(t+1)E$ and $((t+1)E)^\prime$ are the same, $\frac{1}{ |\emph{D}_{k,E}|} \sum_{\left(x_{k,i},y_{k,i}\right) \in \emph{D}_{k,E}}^{|\emph{D}_{k,E}|} \ell_d^{((t+1)E)^\prime}(f^+\left(\omega_k ; x_{k,i}\right),y_{k,i})$ and $\frac{1}{ |\emph{D}_{k,E}|} \sum_{\left(x_{k,i},y_{k,i}\right) \in \emph{D}_{k,E}}^{|\emph{D}_{k,E}|} \ell_d^{(t+1)E}(f^+\left(\omega_k ; x_{k,i}\right),y_{k,i})$ are equal, then we get $(s6)$. Based on  $\|a-b\|_2-\|a-c\|_2 \leq\|b-c\|_2$, $(s7)$ is obtained. 
Since  FedRFQ has embedded a quality detection mechanism in the consensus algorithm, $\mathcal{P}^{t+2}$ is the global prototype of the remaining prototypes after filtering out some of the prototypes. We set $\mathcal{P}^{\prime(t+2)}$ to be the aggregation result without filtering prototypes, so we have $\mathcal{P}^{t+2}$= $\mathcal{P}^{\prime(t+2)}$ - $\mathtt{diff^{t+2}}$, where $\mathtt{diff^{t+2}}$ denotes the gap caused by the quality detection mechanism at the ($t+2$)-th round. Similarly, $\mathtt{diff^{t+1}}$ is the gap at the ($t+1$)-th round, thus we get $(s99)$.
According to Eq. (\ref{eq9}) that computes the global prototype based on the local prototypes, we have $\mathcal{P}=\sum_{k=1}^{K}r_k \bar{P}_{k}$, where $r_k$ denotes the weight in aggregation, so we can get $(s8)$. In our hypothesis, malicious clients adopt an untargeted attack to hide their malicious behaviors, which results in a change in the values of some prototypes. However, this change in the prototype values is in the normal range (if it is not in the normal range, it cannot be successfully disguised as an honest client). Thus, we get the range constraint for $\mathtt{diff^{t}}$:$|\mathtt{diff^{t}}|_2 \leq \varepsilon$,
where $\varepsilon$ is a sufficiently small positive value to indicate the maximum range of normal variations. Similarly, we get $\left|\mathtt{diff^{t+1}}\right|_2 \leq \varepsilon$ and $\left|\mathtt{diff}^{t+2}\right|_2 \leq \varepsilon$. According to the triangular inequality, we can get $\left|\mathtt{diff}^{t+1} - \mathtt{diff}^{t+2}\right|_2 \leq \left|\mathtt{diff^{t+1}}\right|_2 + \left|\mathtt{diff^{t+2}}\right|_2 \leq 2\varepsilon$, and $(s9)$ is based on Eq. (\ref{ee1}) and Eq. (\ref{pingjun}). Then,  we simply weight $r_k \frac{1}{\left|D_k\right|} \sum_{i=1}^{\left|D_k\right|}\mathtt{softpool}\left(f_k\left(\phi_{k}^{(t+1)E};x_{k,i}\right)\right)$ as $n_k \mathtt{softpool}\left(f_k\left(\phi_{k}^{(t+1)E};x_{k}\right)\right)$, we can get $(s10)$. Based on $\|\sum a_i\|_{2} \leq \sum \|a_i\|_{2}$, $(s11)$ can be obtained. As the SoftPool pooling process satisfies $L_2^\prime$-Lipschitz continuity, the proof process is as in Appendix \ref{Proof111},  we have $(s12)$. Further, $(s13)$ is also based the \textit{Assumption} \ref{assumption4}. According to the gradient update regulation, we have $\omega^{(t+1)E}=\omega^{tE}+\eta\|\sum_{e=1}^{E-1}g^{tE+e}+g^{(tE)^\prime}\|_{2}$, so we get ($s14)$. And $(s15)$ is obtained based on $\left\|\sum a_i\right\|_2 \leq \sum\left\|a_i\right\|_2$.

Take expectations of random variable $x_k$ on both sides, we obtain $(s16)$ in Eq. (\ref{222}). Then, based on the \textit{Assumption} \ref{assumption3}, $(s17)$ is obtained.
\begin{footnotesize}
	\begin{equation}
		\begin{aligned}
			\bm{\ell}^{((t+1)E)^\prime} & =\bm{\ell}^{(t+1) E}+\bm{\ell}^{((t+1)E)^\prime}-\bm{\ell}^{(t+1)E} \\
   & \stackrel{(s5)}{=} \bm{\ell}^{(t+1)E}+ \frac{1}{ |\emph{D}_{k,E}|} \sum_{\left(x_{k,i},y_{k,i}\right) \in \emph{D}_{k,E}}^{|\emph{D}_{k,E}|} \ell_d^{((t+1)E)^\prime}(f^+\left(\omega_k ; x_{k,i}\right),y_{k,i}) + \lambda\left\|P^{t+2}-\mathcal{P}^{t+2} \right\|_{2} \\
   &-\frac{1}{ |\emph{D}_{k,E}|} \sum_{\left(x_{k,i},y_{k,i}\right) \in \emph{D}_{k,E}}^{|\emph{D}_{k,E}|} \ell_d^{(t+1)E}(f^+\left(\omega_k ; x_{k,i}\right),y_{k,i})-\lambda\left\|P^{t+2}-\mathcal{P}^{t+1} \right\|_{2}\\
			&\stackrel{(s6)}{=} \bm{\ell}^{(t+1)E}+ \lambda\left\|P^{t+2}-\mathcal{P}^{t+2} \right\|_{2}-\lambda\left\|P^{t+2}-\mathcal{P}^{t+1} \right\|_{2}\\
			& \stackrel{(s7)}{\leq} \bm{\ell}^{(t+1)E}+\lambda\left\|\mathcal{P}^{t+2}-\mathcal{P}^{t+1}\right\|_2 \\
			& \stackrel{(s99)}{=} \bm{\ell}^{(t+1)E}+\lambda\left\|(\mathcal{P}^{\prime(t+2)}-\mathsf{diff^{t+2}})-(\mathcal{P}^{\prime(t+1)}-\mathsf{diff^{t+1}})\right\|_2 \\
			& \stackrel{(s8)}{=} \bm{\ell}^{(t+1)E}+\lambda\left\|\sum_{k=1}^K r_k \bar{P}_{k}^{(t+1)E}-\sum_{k=1}^K r_k \bar{P}_{k}^{tE} \right\|_2+\left\|\mathsf{diff^{t+1}}-\mathsf{diff^{t+2}}\right\|_2 \\
			& \stackrel{(s9)}{\leq} \bm{\ell}^{(t+1)E}+\lambda \| \sum_{k=1}^K r_k \frac{1}{\left|\emph{D}_k\right|} \sum_{i=1}^{\left|\emph{D}_k\right|}\mathtt{softpool}\left(f_k\left(\phi_{k}^{(t+1)E};x_{k,i}\right)\right)-\sum_{k=1}^K r_k \frac{1}{\left|\emph{D}_k\right|} \sum_{i=1}^{\left|\emph{D}_k\right|}\mathtt{softpool}\left(f_k\left(\phi_{k}^{tE} ; x_{k,i}\right)\right) \|_2+ 2\varepsilon\\
			& \stackrel{(s10)}{=} \bm{\ell}^{(t+1)E}+\lambda \|\sum_{k=1}^K n_k \mathtt{softpool}\left(f_k\left(\phi_{k}^{(t+1)E};x_{k}\right)\right)- \sum_{k=1}^K n_k \mathtt{softpool}\left(f_k\left(\phi_{k}^{tE} ; x_{k}\right)\right) \|_2 +2\varepsilon\\
			& \stackrel{(s11)}{\leq} \bm{\ell}^{(t+1)E}+\lambda \sum_{k=1}^K n_k \|\mathtt{softpool}\left(f_k\left(\phi_{k}^{(t+1)E}; x_{k}\right)\right)-\mathtt{softpool}\left(f_k\left(\phi_{k}^{tE} ; x_{k}\right)\right) \|_2+2\varepsilon \\
			& \stackrel{(s12)}{\leq} \bm{\ell}^{(t+1)E}+\lambda L_{2}^\prime \sum_{k=1}^K n_k  \| f_k\left(\phi_{k}^{(t+1)E}; x_{k}\right)-f_k\left(\phi_{k}^{tE} ; x_{k}\right) \|_2 +2\varepsilon\\
			& \stackrel{(s13)}{\leq} \bm{\ell}^{(t+1)E}+\lambda L_{2} L_{2}^{\prime} \sum_{k=1}^K n_k  \| \omega_{k}^{(t+1)E}-\omega_{k}^{tE} \|_2 +2\varepsilon\\
			& \stackrel{(s14)}{=} \bm{\ell}^{(t+1)E}+\lambda L_{2} L_{2}^{\prime}\eta \sum_{k=1}^K n_k  \| \sum_{e=1}^{E-1} g_{k}^{tE+e}+ g_{k}^{(tE)^\prime}\|_2+2\varepsilon\\
			& \stackrel{(s15)}{\leq} \bm{\ell}^{(t+1)E}+\lambda L_{2} L_{2}^{\prime} \eta \sum_{k=1}^K n_k (\sum_{e=1}^{E-1} \|  g_{k}^{tE+e} \|_2 + \|  g_{k}^{(tE)\prime} \|_2)+2\varepsilon. \\
			\label{21}
		\end{aligned}
	\end{equation}
\end{footnotesize}
\begin{footnotesize}
	\begin{equation}
		\begin{aligned}
			\mathbb{E}[\bm{\ell}^{((t+1)E)^\prime}] &\stackrel{(s16)}{\leq} \bm{\ell}^{(t+1)E}+\lambda L_{2} L_{2}^{\prime} \eta \sum_{k=1}^K n_k (\sum_{e=1}^{E-1} \mathbb{E}[\| g_{k}^{tE+e} \|_2]+\mathbb{E}[\| g_{k}^{(tE)^\prime} \|_2]) + 2\varepsilon \\
			&\stackrel{(s17)}{\leq} \bm{\ell}^{(t+1)E}+\lambda L_{2} L_{2}^{\prime} \eta E G + 2\varepsilon. \\ 
			\label{222}
		\end{aligned}
	\end{equation}
 \end{footnotesize}

\end{IEEEproof}

\textbf{Proof of Theorem \ref{theorem1}.}
\begin{IEEEproof}
Since \textit{Lemma} \ref{lemma1} and \ref{lemma2} hold, for an arbitrary client $k$,Theorem \ref{theorem1} holds in Eq. (\ref{theormaa}) after each round.
 \begin{footnotesize}
	\begin{equation}
			\mathbb{E}[\bm{\ell}_k^{((t+1)E)^\prime}]\leq \bm{\ell}_k^{(t E)^\prime}-\left(\eta-\frac{L_a \eta^2}{2}\right) (\sum_{e=1}^{E-1}\left\|\nabla \bm{\ell}_k^{tE+e}\right\|_2^2+\left\|\nabla \bm{\ell}_k^{(tE)^\prime}\right\|_2^2)
			+\frac{L_a E \eta^2}{2} \sigma^2 +\lambda L_{2} L_{2}^{\prime} \eta E G + 2\varepsilon.
			\label{theormaa}
		\end{equation}
  \end{footnotesize}
	
\end{IEEEproof}
\section{The SoftPool  Satisfies  $L^\prime_2$-Lipschitz Continuity}\label{Proof111}
\begin{IEEEproof} 
Firstly, we combine Eq. ($\ref{e6}$) and Eq. ($\ref{ee12}$) to get Eq. ($\ref{ee6}$), and we verify that the following Eq. (\ref{ee6}) satisfies $L^\prime_2$-Lipschitz continuity.
\begin{footnotesize}
\begin{equation}
\tilde{P}(C_{k,i}^\mathbf{R}) = \sum_{i \in \mathbf{R}} \frac{\exp(C_{k,i}^{\mathbf{R}_i})}{\sum_{v \in \mathbf{R}} \exp(C_{k,i}^{\mathbf{R}_v})} C_{k,i}^{\mathbf{R}_i} \label{ee6}.
\end{equation}
\end{footnotesize}

We need to prove that for any $ C_{k,i}^\mathbf{R}$ and $C_{k,i}^{\mathbf{R}^{\prime}}$, there exists a constant $L^\prime_2$ such that the following equation holds. 
\begin{footnotesize}
\begin{equation}
	\left|\tilde{P}\left(C_{k,i}^\mathbf{R}\right)-\tilde{P}\left(C_{k,i}^{\mathbf{R}^{\prime}}\right)\right| \leq L^\prime_2\left|C_{k,i}^\mathbf{R}-C_{k,i}^{\mathbf{R}^{\prime}}\right|.
 \nonumber
\end{equation}
\end{footnotesize}

Because of $\sum_{v \in \mathbf{R}} \exp(C_{k,i}^{\mathbf{R}_v}) \neq 0$, it can avoid the case where the denominator is $0$. Next, we calculate $\tilde{P}\left(C_{k,i}^\mathbf{R}\right)$ and $\tilde{P}\left(C_{k,i}^{\mathbf{R}^{\prime}}\right)$.
\begin{footnotesize}
\begin{equation}
	\begin{aligned}
		& \tilde{P}\left(C_{k,i}^\mathbf{R}\right)=\sum_{i \in \mathbf{R}} \frac{\exp(C_{k,i}^{\mathbf{R}_i})}{\sum_{v \in \mathbf{R}} \exp(C_{k,i}^{\mathbf{R}_v})} C^{\mathbf{R}_i}, \\
		& \tilde{P}\left(C_{k,i}^{\mathbf{R}^{\prime}}\right)=\sum_{i \in \mathbf{R}^{\prime}} \frac{\exp(C_{k,i}^{\mathbf{R}^{\prime}_i})}{\sum_{v \in \mathbf{R}^{\prime}} \exp(C_{k,i}^{\mathbf{R}^{\prime}_v})} C_{k,i}^{\mathbf{R}^{\prime}_i}.
	\end{aligned}
 \nonumber
\end{equation}
\end{footnotesize}

Then, we compute $ \left|\tilde{P}\left(C_{k,i}^\mathbf{R}\right)- \tilde{P}\left(C_{k,i}^{\mathbf{R}^{\prime}}\right)\right|$ in Eq. (\ref{ees}), where $(s18)$ follows from $0 \leq \frac{\exp(C_{k,i}^{\mathbf{R}_i})}{\sum_{v \in \mathbf{R}} \exp(C_{k,i}^{\mathbf{R}_v})} \leq 1$, $(s19)$ follows from the fact that $\mathbf{R}$ and $\mathbf{R}^\prime$ have kernel regions $\hat{k}^2$ of the same size, $(s20)$ follows from $\left|\sum a \right| \leq \sum \left| a \right|$, $(s21)$ follows from the fact that  there always exists a ($\hat{k}^2-1$) when $\left|C_{k,i}^{\mathbf{R}}- C_{k,i}^{\mathbf{R}^{\prime}}\right|_{\mathsf{max}}$ denotes the maximum absolute value of the difference between the elements in the same position of $\mathbf{R}$ and $\mathbf{R}^{\prime}$.
\begin{footnotesize}
\begin{equation}
	\begin{aligned}			    & \left|\tilde{P}\left(C_{k,i}^\mathbf{R}\right)- \tilde{P}\left(C_{k,i}^{\mathbf{R}^{\prime}}\right)\right| \\
		&=\left|\sum_{i \in \mathbf{R}} \frac{\exp(C_{k,i}^{\mathbf{R}_i})}{\sum_{v \in \mathbf{R}} \exp(C_{k,i}^{\mathbf{R}_v})} C_{k,i}^{\mathbf{R}_i}-\sum_{i \in \mathbf{R}^{\prime}} \frac{\exp(C_{k,i}^{\mathbf{R}^{\prime}_i})}{\sum_{v \in \mathbf{R}^{\prime}} \exp(C_{k,i}^{\mathbf{R}^{\prime}_v})} C_{k,i}^{\mathbf{R}^{\prime}_i}\right| \\
		&\stackrel{(s18)}{\leq} \left|\sum_{i \in \mathbf{R}} C_{k,i}^{\mathbf{R}_i}-\sum_{i \in \mathbf{R}^{\prime}}  C_{k,i}^{\mathbf{R}^{\prime}_i}\right| \\
		&\stackrel{(s19)}{=} \left|\sum^{\hat{k}^2-1}_{i=0} C_{k,i}^{\mathbf{R}_i}-\sum^{\hat{k}^2-1}_{i=0}  C_{k,i}^{\mathbf{R}^{\prime}_i}\right| \\
		&\stackrel{}{=} \left|\sum^{\hat{k}^2-1}_{i=0} (C_{k,i}^{\mathbf{R}_i}-C_{k,i}^{\mathbf{R}^{\prime}_i})\right| \\
		&\stackrel{(s20)}{\leq} \sum^{\hat{k}^2-1}_{i=0} \left|C_{k,i}^{\mathbf{R}_i}- C_{k,i}^{\mathbf{R}^{\prime}_i}\right| \\
		&\stackrel{(s21)}{\leq}  (\hat{k}^2-1)\left|C_{k,i}^{\mathbf{R}}- C_{k,i}^{\mathbf{R}^{\prime}}\right|_{\mathsf{max}}.\label{ees}
    \end{aligned}
\end{equation}
\end{footnotesize}

Therefore, the SoftPool can get $(\hat{k}^2-1)$ that is regarded as $L^\prime_2$ to satisfy $L^\prime_2$-Lipschitz continuity in Eq. (\ref{ee5}).
\begin{footnotesize}
\begin{equation}
	     \left|\tilde{P}\left(C_{k,i}^\mathbf{R}\right)- \tilde{P}\left(C_{k,i}^{\mathbf{R}^{\prime}}\right)\right| \\
		\leq  (\hat{k}^2-1)\left|C_{k,i}^{\mathbf{R}}- C_{k,i}^{\mathbf{R}^{\prime}}\right|_{\mathsf{max}}.
\label{ee5}
\end{equation}
\end{footnotesize}
\end{IEEEproof}
\end{appendices}
\end{document}